**Interface study of thermally driven chemical kinetics involved in Ti/$Si_3N_4$ based metal-substrate assembly by X-ray photoelectron spectroscopy**

Sachin Yadav[a,b] and Sangeeta Sahoo[a,b,*]

[a]*Academy of Scientific and Innovative Research (AcSIR), AcSIR Headquarters CSIR-HRDC Campus, Ghaziabad, Uttar Pradesh 201002, India.*

[b]*Electrical & Electronics Metrology Division, National Physical Laboratory, Council of Scientific and Industrial Research, Dr. K. S. Krishnan Marg, New Delhi-110012, India.*

[*]Corresponding author.

*E-mail address:* sahoos@nplindia.org (Sangeeta Sahoo)

**Abstract**

Diffusion mediated interaction in metal-substrate assembly during high temperature annealing leads to possible formation of new composite materials. Here, sputtered grown Ti films on $Si_3N_4$/Si substrate has been reported to produce titanium nitride and silicide based binary composites while undergoing high vacuum annealing process at temperatures 650°C and above. Diffusion of thermally decomposed Si and N atoms from $Si_3N_4$ and their subsequent chemical reaction with Ti have been probed by X-ray photo electron spectroscopy. For annealing at 800°C and above, most of the Si atoms show preferences to stay in elemental form rather than developing silicide phase with Ti. Whereas at lower annealing temperature, silicide becomes the dominant phase for decomposed Si atoms. However, N atoms react promptly with Ti and form TiN which appears as the majority phase for each of the studied annealing temperature. Further, the nitride and silicide phases across the films have been compared quantitatively for various annealing temperature and the maximum silicide formation is observed for the sample annealed at 780°C. Finally, the thermally anchored metal-substrate interaction mechanism can be exploited to fabricate disordered superconducting TiN films where $TiSi_2$ and Si can be used to tune the level of disorder by altering the annealing temperature.

## 1. Introduction

Owing to their thermal stability and low electrical resistivity, transition metal silicide and nitrides are widely used as source, drain, gate electrodes and electrical interconnects in complementary- metal-oxide-semiconductor (CMOS) compatible devices [1-3]. Commonly, high temperature annealing of metallic thin film surrounded by Si or $SiO_2$ is adhered for the fabrication of low resistive metal silicide [4, 5]. However, while forming diffusion mediated silicide by thermally anchored metal and silicon atoms, lateral diffusion often causes short circuit by making a connecting electrical path through the spacer layer separating the electrodes. To address this issue, annealing in presence of $N_2$ ambient and use of nitrogen-based substrate like $Si_3N_4$ have been introduced [4, 6]. Further, it has been reported that TiN can act like diffusion barrier to prevent the formation of laterally diffused silicide [7]. Therefore, a combination of silicide and nitride of a common transition metal could perhaps serve as an ideal candidate to achieve low resistive interconnects with no electrical connection between the electrodes. Recently, the combination of $TiSi_2$/TiN has been used as full metal gate in dual-channel CMOS technology [8].

In this work, we present a substrate mediated technique which can form nitrides and silicides simultaneously for the transition metal, Ti, by high temperature annealing of Ti/$Si_3N_4$ based metal-substrate assembly under high vacuum condition. Ti is known to react with conventional oxide and nitride-based substrates due to its reactive nature towards elements like oxygen, nitrogen and silicon under certain conditions like exposure to air, thermal treatment etc. [3, 6, 9]. As the substrates like $SiO_2$ and/or $Si_3N_4$ contain two elements that can interact simultaneously with Ti, it is essential to know how Ti reacts with both the elements at the same time. For example, TiN and $TiSi_2$ based binary stable phases are expected to form when Ti/$Si_3N_4$ assembly is undergone a high temperature annealing treatment [6, 10] but at the same time, a clear understanding is needed about how these two binary phases compete with each other and how the annealing temperature ($T_a$) can influence the competition. Therefore, a thorough quantitative analysis might be helpful to understand and determine the binary phases and their evolution with $T_a$ across the film and the interface.

While understanding the interaction between metallic Ti and the underneath $Si_3N_4$ substrate, one can use the high temperature annealing as an effective technique to fabricate the stable binary TiN and $TiSi_2$ which further can contribute to the formation of ternary Ti-Si-N based composite materials having promising optoelectronic properties [11]. Depending on the external parameters like annealing temperature, film thickness, pressure etc., the preferential growth may occur between the nitride and silicide phases [6] and the majority phase dominates strongly over the minority so that distinctive characteristic properties from the former can be explored while keeping the latter unaffected. For instance, some of the characteristic properties, such as superconductivity [12, 13], plasmonic [14-16], bio-compatibility [17] and coating [18, 19] related properties that are possessed by TiN, can be explored if it is obtained as the majority phase governing the properties of the whole composite system.

Here, we have probed the metal-substrate interaction in Ti/$Si_3$N4 assembly with respect to annealing temperature ($T_a$) which plays an important role as it leads to the decomposition of $Si_3N_4$ into elemental N and Si, and simultaneously, it provides thermal energy for the decomposed elements to diffuse inside the metal [20, 21]. The $T_a$ has been varied in the range from 650°C to 820°C. The diffusion of the decomposed elements and their subsequent interaction with Ti for the formation of nitride and silicides are confirmed and have been probed by high resolution X-ray photo electron spectroscopy (XPS). We have observed that the amount of decomposed N and Si atoms and their interaction with Ti strongly depends on $T_a$. This is obvious that the decomposition process for $Si_3N_4$ gets boosted with increasing $T_a$, however, the diffusion of the decomposed N and Si atoms inside the metal and their subsequent interaction with Ti show anomalous behaviour with $T_a$. For example, we observe that the amount of Si atoms present in the metal side reaches its maximum value at $T_a$= 780°C and the same decreases with further increasing $T_a$. Accordingly, the maximum amount of silicide formation is observed for $T_a$= 780°C. Interestingly further, depending on the $T_a$, the decomposed Si atoms, inside the metallic film, show preferences towards either the formation of silicide phases or to be remained in its elemental form. For $T_a \geq$ 800°C, majority of the decomposed Si atoms prefer to stay in their elemental form with very little/negligible contribution towards the silicide formation. Hence, in this range of $T_a$, the annealing leads to the fabrication of mainly TiN embedded with elemental Si and the minority phase $TiSi_2$. Further, a quantitative analysis of the two product phases, namely, TiN and $TiSi_2$, has been carried out for better understanding of reaction kinetics of Ti with N and Si across the film and at the interface. Here, one of the key observations is that the appearance of TiN as the majority phase for each of the studied $T_a$ unambiguously. Therefore, this thermally driven metal-substrate interaction technique can be used as an efficient nitridation technique, which is different than the other conventional methods requiring external source of nitrogen [18, 22-26] to produce TiN. Of particular interest towards the superconducting properties, ultrathin films of disordered TiN have exhibited various fundamental quantum phenomena such as superconductor-insulator (SIT) quantum phase transition [27], quantum criticality [28], quantum phase slip etc. [29]. Here interestingly, the majority TiN phase, inherently doped with elemental Si and $TiSi_2$, can serve as a model system to study disordered superconductivity and the related quantum phenomena where the latter two can play the role of disorder. *It is noteworthy to mention that we employ the presented technique to prepare disordered superconducting ultrathin films of TiN to explore the aforesaid quantum phenomena and our preliminary results demonstrate that the superconducting TiN films produced by this method can compete with the reported high* *quality TIN films prepared by other conventional techniques and the results will be published elsewhere.*

## 2. Experimental

### 2.1 Synthesis process

We employed intrinsic undoped Si(100) wafer topped with low pressure chemical vapor deposition (LPCVD) grown $Si_3N_4$ dielectric spacer layer of 80 nm thickness as the substrate on which Ti films were grown by DC magnetron sputtering. At first, the substrates were cleaned by using the standard cleaning process involving sonication in acetone and isopropanol bath for 15 minutes each. Thereafter, the cleaned samples were loaded and transferred into the ultra-high vacuum (UHV) chamber of the sputtering unit where they were pre-heated at about 820 ˚C for 30 minutes under very high vacuum condition ( $p \sim 5 \times 10^{-8}$ Torr). The pre-heating was performed to remove any adsorbed or trapped molecules/residue on the surface of the substrate. After performing all the cleaning processes, cleaned $Si_3N_4$/Si substrate was cooled down to room temperature and had been transferred to the sputtering chamber without breaking the vacuum. A thin layer of Ti with thickness of about 15 nm was then deposited on the substrate by dc magnetron sputtering of Ti (99.995% purity) in the presence of high purity Ar (99.9999%) gas with a rate of 5 nm/min. Sputtering of Ti was performed with a base pressure less than $1.5 \times 10^{-7}$ Torr. Finally, the sputtered sample was transferred again into the UHV chamber for annealing. The various assemblies of Ti films on $Si_3N_4$/Si (100), shown in Fig. 1(a), were then annealed at temperatures ranging between 650°C and 820°C for 2 hrs. at pressure less than $5 \times 10^{-8}$ Torr. The temperature variations during the annealing process are graphically shown in Fig. 1(b). For the present study, we fabricated thin films on substrates with an area of 5 mm x 5 mm.

## 2.2 Structural characterizations

Structural characterizations by grazing incidence X-ray diffraction (GIXRD) were done by PAnalytical PRO MRD X'pert X-ray diffractometer with CuKα radiation operating at 40KV and 30 mA. X-ray Photoelectron Spectroscopic (XPS) measurements were performed in UHV based fully integrated XPS instrument, Thermo Scientific Nexsa, operating at a base pressure of $1.5 \times 10^{-7}$ pascals. A monochromatic, micro-focused, low power (75 Watt) Al-Kα X-ray source (1486.7 eV) was employed with radiation spot size of 400 μm for the excitation. A 180° double focusing hemispherical analyser, equipped with a 128-channel detector, was used to collect the exited photo electrons. To study the depth profile, minimizing the contribution of native oxides & contaminants, and to probe the interface, in-situ sputtering via energetic $Ar^+$ ions (at 500 eV) was performed with fixed interval of time (15 sec). The XPS spectra were collected at an analyser pass energy of 150 eV in depth profile operational mode known as "snapshot mode". By increasing the analyser pass energy from 50 eV to 150 eV in the snapshot mode, one reduces the spectra acquisition time greatly but at the expense of spectra resolution. In this mode, electron signal from entire range can be collected simultaneously by the attached128 channel detectors of the analyser and a maximum number of XPS scans related to the steps in depth profile can be performed simultaneously, hence the acquisition time would be greatly reduced [30]. Low energy electrons and ions from a dual beam charge neutralization system were used to neutralize the excess charge. The adventitious C-C bond at 284.8 eV was considered for the referencing of binding energy. Avantage Software was used for instrument control, data acquisition, data processing, and data analysis.

## 3. Results and Discussions

The process of interaction between Ti thin film and $Si_3N_4$ substrate under high temperature annealing and the formation of substrate mediated nitride and silicide phases of Ti have been illustrated schematically in Fig. 1. First, we deposited Ti thin films by DC magnetron sputtering on $Si_3N_4$/Si substrate, shown in Fig. 1(a), and thereafter the samples were transferred in situ to an attached UHV chamber to undergo a high temperature annealing process [Fig. 1(b)]. The temperature variation with time during the annealing process is shown by color-gradient lines in Fig. 1(b), where the reddish shaded rectangular block represents the range of annealing temperature used for the present study. Here, the annealing temperature is varied in the range between 650°C to 820°C. At temperature of about 650°C and above, we have observed that $Si_3N_4$ layer starts to decompose into elemental silicon and nitrogen [21] which then move into Ti film by thermally anchored diffusion process [31]. At the interface of Ti/ $Si_3N_4$, Ti atoms may also diffuse into the substrate. The diffusion mediated metal-substrate interaction across the interface during the annealing is illustrated schematically in Fig. 1(c). Due to high affinity of titanium towards both nitrogen and silicon, titanium nitride and silicide phases are formed at higher temperature [31, 32]. However, the lighter N atoms react fast with titanium and form the stable stoichiometric cubic TiN as the majority phase. While, relatively lesser number of available Si atoms from $Si_3N_4$ and their heavier mass compared to that of N atoms make the silicide phase as the minority phase. Further depending on $T_a$, it has been observed that decomposed Si atoms partly stay in their elemental form instead of taking part into the silicide formation. This might be due to the unavailability of Ti to form silicide after the formation of dominant TiN stable phase or due to a strong dependence of silicide formation on the annealing temperature and also due to the unstable nature of silicide in presence of $Si_3N_4$ [6]. Besides, the relatively heavier Si atoms stay close to the interface of $Si_3N_4$/Ti and may diffuse into the Si substrate through the interface of $Si_3N_4$/Si. The transformation of Ti film into a composite film of TiN and $TiSi_2$ is shown in Fig. 1(d). Here, annealing temperature plays the most significant role on all the three main mechanisms, *viz.* (i) decomposition of $Si_3N_4$ into Si and N atoms, (ii) diffusion of the decomposed atoms and (iii) governing the chemical reaction involved in the present scenario.

$$Si_3N_4 \xrightarrow{heating\ (\geq 650°C)} 3Si + 2N_2 \quad (1)$$

$$\Delta G_T^{\circ}(kJ.mole^{-1}) = 744 - T(K) * (0.327)$$

$$Ti(s) + \frac{1}{2}N_2 = TiN(s) \quad (2)$$

$$\Delta G_T^{\circ}(kJ.mole^{-1}) = -338 + T(K) * (0.09633)$$

$$Ti(s) + 2Si(s) = TiSi_2(s) \quad (3)$$

$$\Delta G_T^{\circ}(kJ.mole^{-1}) = -134 + T(K) * (0.0074)$$

$$5Ti(s) + Si_3N_4(s) = 4TiN(s) + TiSi_2(s) + Si(s) \quad (4)$$

$$\Delta G_T^{\circ}(kJ.mole^{-1}) = -740 + T(K) * (0.0668)$$

The chemical reactions that take place during the annealing process are shown in Equations (1-4) along with the Gibbs's free energy expressions at temperature *T(K)* [33, 34]. The Gibbs's free energy for the decomposition of $Si_3N_4$ into Si and N [Equation (1)] indicates that the reaction is not favored for the temperature range considered in this study. However, impurities play a major role to promote the decomposition process at much lower temperature than that is thermodynamically required to obtain the condition, $\Delta G_T^{\circ} < 0$ [35, 36]. During the course of sample processing in practice, the impurities are introduced inevitably and particularly, the impurities like carbon, oxygen can be present in most of the cases. It has been shown that the presence of carbon actually boosts the decomposition by breaking Si-N bond and forming SiC and N [37]. Besides, metallic elements like Fe, Ti etc. are known to influence the decomposition process greatly because of their reactive nature [35]. The reported observations on the interaction between Ti and $Si_3N_4$ during annealing for a wide range of temperature clearly indicate a major role of Ti in decomposition of $Si_3N_4$ at moderate temperature range starting from 400°C [6, 32, 38]. The chemical reaction between Ti and $Si_3N_4$ is expressed in Equation (4) which is thermodynamically spontaneous reaction for the range of temperature usually used in the laboratory-based experiments. As seen from Equation (4), the presence of Si does not have much influence on the reaction process and its thermodynamics, however, at higher temperature, the presence of excess Si is observed. Further, from the Gibb's free energy values corresponding to Equations (2) and (3), TiN appears to be more stable than $TiSi_2$ [38, 39]. Finally, from Equations (2), (3) and (5) it is clear that TiN is the dominant majority phase among the nitride and silicide phases.

In Fig. 2, we have displayed the X-ray diffraction (XRD) pattern obtained from three representative samples. Two of them were annealed at 820°C and differ in thickness (25 nm & 15 nm) as represented by the bottom two spectra in Fig. 2, while, the top spectrum represents a 15 nm thick film annealed at 800°C. Strong peaks related to stable cubic stoichiometric TiN appear along with a couple of substrate peaks. However, no traceable peaks from silicide phase has been observed. Further, the peaks are more prominent for the bottom two samples that were annealed at 820°C compared to the top one which was annealed at 800°C. With further lowering the annealing temperature ($T_a$), we find difficulties to resolve even nitride related peaks. Therefore, it is confirmed from the XRD characterization that the nitridation happened via the decomposition of Si3N4 during the annealing process as there was no external source of nitrogen except for the ones present in the $Si_3N_4$ substrate. Hence, Ti is transformed into TiN by following the annealing process as explained in Fig. 1. However, we have obtained a very limited information from the XRD data about the chemical reactions occurring for lower annealing temperature. Further, XRD analysis does not provide any insight into the formation of silicide phase which in principle should also be formed. In order to have clear understanding of the effect of annealing temperature on the final outcome like the nitride and silicide phases, we have carried out X-ray photoelectron spectroscopy (XPS) analysis on the samples those were annealed at different temperatures ranging between 650°C and 820°C. In the following section, we have presented an in-depth XPS analysis performed on various samples with varying $T_a$.

In order to have clear understanding on the chemical kinetics involved in the formation of nitride and silicide phases, we have performed interface studies on annealed thin film samples by using X-ray photoelectron spectroscopy (XPS) technique. The elemental core level binding energy (XPS) spectra for a representative reference sample annealed at 820°C and of thickness 15 nm have been shown in Fig. 3 which includes a set of Ti 2p, N 1s, Si 2p and O 1s core level spectra in (a), (b), (c) &(d), respectively. The depth profile is obtained by etching the film with $Ar^+$ ions with energy of 500 eV in 15 sec time intervals from the top surface till it reached to the substrate with subsequent XPS scans recorded after each etching step.

In Fig. 3, all the relevant XPS scans, recorded after every 30 sec etching intervals, are placed as a set of spectra that are shifted in upward direction for clarity. The direction from the top surface towards the substrate is shown by the black vertical arrow in Fig. 3(a). The most common peaks related to nitride, silicide and oxides are shown by dotted lines in the core level spectra of abovementioned four elements presented in Fig. 3. The core level spectra of Ti 2p, as presented in Fig. 3 (a), display two prominent peaks related to its spin-orbit split doublets, Ti $2p_{3/2}$ and Ti $2p_{1/2}$, at 454.9 eV and 460.7 eV, respectively. The binding energy position for Ti $2p_{3/2}$ at 454.9 eV and the doublets spacing of 5.8 eV refer to stoichiometric TiN[40]. No other prominent peak appears in the Ti 2p spectra except for some broad features. However, by deconvolution of spectra, the presence of any other possible chemical states of Ti can be resolved and the same is discussed later in the article. The binding energy position of other common chemical states of Ti, namely, the metallic Ti ($Ti^0$) [41], $TiSi_2$ [42, 43], oxides ($Ti^{2+}$, $Ti^{3+}$ & $Ti^{4+}$) [41, 42] and oxynitride (Ti-N-O) [44, 45] are shown in Fig. 3(a). It is evident that the dominant contribution comes from TiN. Further, the XPS scans, recorded after about 360 sec of etching from the interface, exhibit gradual shifting of peaks towards higher binding energy. Here the peak intensity is very low and the peaks shift towards the binding energy of titanium oxynitride which can be formed inside the substrate by diffused Ti and any oxygen atoms at the interface.

The formation of stoichiometric TiN is confirmed by the presence of a strong peak appearing at 397 eV for N 1s spectra as shown in Fig. 3(b)[46]. Here, the $Ar^+$ ion sputtering time between two consecutive spectra is 30 sec as mentioned in Fig. 3(a). Now, as we move towards the substrate through the film, after 360 sec of etching, the presence of $Si_3N_4$ from the substrate becomes evident at ~ 398 eV [47, 48] in N 1s. As we keep on etching to reach further down to the substrate, the peak shifts towards higher binding energy and after 600 sec of sputtering the peak shifts to 400.2 eV which can be assigned to a possible oxynitride (N-O) bonding. Further the peak shifting in the intermediate binding energy region between $Si_3N_4$ and N-O can be related to Si oxynitride (Si-N-O) with compositional variations. Actually, the shifting towards higher binding energy is observed for all the four elements, i.e. Ti 2p, N 1s, Si 2p and O 1s, presented in Fig. 3. These shifted higher energy peaks appear more prominent in N 1s and Si 2p spectra. For the latter in Fig. 3(c), we observe a broad peak spanning in the range from 98 eV to 101 eV with the center at ~ 99.5 eV which can be assigned to elemental Si[47, 49, 50]. The binding energy position for $TiSi_2$ at 98.8 eV[47] appears on the shoulder of the peak and it is shown in the figure. On the top surface, we observe another broad peak centered at ~ 103 eV in between 101 eV to 105 eV. This broad peak on the surface is obviously related to silicon oxide and sub-oxides that are formed while the samples get exposed to air and the excess Si from decomposed $Si_3N_4$ reacts with the oxygen present in the ambient air[51]. Various oxidation states of Si are marked in the spectra of Si 2p [52, 53]. Here, it is clear that Si prefers to stay in its elemental form and there might be a minor contribution towards $TiSi_2$ formation. The occurrence of $SiO_2$ and other sub oxides on the surface again indicates the presence of elemental Si in the film. It is noteworthy to mention that no strong oxide peak appears on the surface for Ti 2p in Fig. 3(a), which indicates that there might be a very little or no Ti left for taking part into the oxidation when the sample gets exposed to air. Most of the Ti transformed into TiN along with a minority contribution for $TiSi_2$. Therefore, non-availability of Ti might be one of the reasons for Si to stay at its elemental form.

By $Ar^+$ ion sputtering, when we move close to the substrate, the peak related to $Si_3N_4$ appears at ~102.2 eV [51, 54]. Further etching leads to peak shifting towards higher binding energy in a similar fashion as that for the other elements, particularly, the N 1s. These peaks can be originated due to the formation of silicon oxynitride (Si-N-O) phases at high temperature and the continuous shift in binding energy indicates the changes in composition with varying number of bonds for N-Si, Si-O and N-O bonds. For example, the shifting in binding energy from 399.3 eV towards higher binding energy in N 1s spectra can be related to the compositional transition such as $Si_3N$-to-$Si_2NO$-to-$SiNO_2$ [55]. It has been already shown in the literature that depending on N concentration the binding energy shifts for silicon oxynitride (Si-N-O) from the binding energy of $Si_3N_4$ to that of $SiO_2$ [56, 57]. Here the shift extends up to 104.4 eV whereas, 103.3 eV is generally considered as the binding energy of $SiO_2$ [58, 59]. However, the binding energy for $SiO_2$ varies widely as reported in the literature, [47, 54], therefore, 104.4 eV might be related to an oxidation state of Si and we refer this as Si-O bond. Correspondingly, similar to N 1s spectra, the compositional variations with the binding energy shift can be interpreted as $Si_3N_4$-to-Si-N-O-to $SiO_x$ in Si 2p [50, 54, 57, 60].

For O 1s spectra shown in Fig. 3(d), a strong peak at ~ 532 eV appears for the surface scan which can be related to an oxidation state of silicon (SiOx) [47]. Further, the appearance of Ti-N-O peak inside the film at 531.5 eV is evident [61]. The shift towards higher binding energy for the spectra collected from the interface for O1s can be related to the formation of oxynitride phase [53]. Here, the shift in Ti 2p spectra with very low intensity towards the binding energy of titanium oxynitride which can be formed inside the substrate by diffused Ti.

The shift towards higher binding energy at the interface has been observed mostly for the samples annealed at Ta≥ 800°C. As all the four elements, namely, Ti 2p, N1s, Si 2p and O1s, exhibit the shift in higher binding energy side, charging of samples during XPS scan and $Ar^+$ ion sputtering might appear as one of the reasons. However, the similar type of shifting is not observed for lower $T_a$. For example, the sample, annealed at 780°C, was etched further up to ~ 800 sec where Ti concentration reached below 1 atm.%. The atomic profile and N 1s spectra for the sample are shown in Fig. S5 in the supplementary Material. Here, no shifting is observed. Further, we have carried out XPS measurements on highly doped substrate [$Si_3N_4$/ Si (p-type)] where we observed the similar type of shifting only for Ta≥ 800°C. Hence, we can conclude that, charging is not the reason behind the shifting occurred for the samples annealed at or above 800°C. Most likely, the formation and mixing of Si-O and N-O bonds occurs at this range of annealing temperature which may lead to the shifting of the peaks towards the higher binding energy.

Finally, the XPS interface studies, shown in Fig. 3, confirm the formation of stoichiometric TiN along with elemental Si and a possible minority $TiSi_2$ phase by annealing $Ti/Si_3N_4$ assembly at 820°C for 2 hours under high vacuum condition. Moreover, we have observed the similar results for the films annealed at 800°C and the experimental data is shown in Fig. S1 in the Supplementary Material.

In order to observe the effect of annealing temperature on the chemical reactions occurring between Ti thinfilms and the underneath $Si_3N_4$ substrate layer, we have carried out XPS interface studies on samples annealed at few different temperatures. In Fig. 4, we have displayed the core level binding energy spectra for four representative samples annealed at 780°C, 750°C, 700°C and 650°C, respectively. All the four samples were having the same Ti thickness of 15 nm each and the growth conditions were kept as close as possible among the samples. As we emphasize on the formation of nitride and silicide phases originated from the interaction of Ti with thermally decomposed Si and N atoms from $Si_3N_4$ layer, we present the core level binding energy spectra of Ti 2p, N 1s and Si 2p in Fig. 4. The corresponding O 1s spectra have been presented in Fig. S2 in Supplementary Material. The rows represent different samples and the columns represent different elements as mentioned in the figure.

Unlike the results presented in Fig. 3, here the as-loaded samples, due to air exposure before loading for the XPS characterization, show strong titanium oxide peak ($TiO_2$) at 458.8 eV [47, 48, 62] by the top surface scan at the bottom of the Ti 2p spectra in Fig. 4. For the sample annealed at 780°C, a broad and prominent peak at 102.8 eV related to Si-O bonding [47] appears in Si 2p spectra for the top surface scan, as presented in Figure 4(c). With reduction in annealing temperature, the peak starts to get flattened at $T_a$ = 750°C and finally it disappears at $T_a$ = 650°C. The top surface scan for Ti 2p indicates the presence of residual Ti on the surface and the same for Si 2p shows a detrimental trend of excess Si present on the film surface with annealing temperature. In order to study the depth profile and interface properties, the samples were etched from the top surface to the interface until it reached to the substrate in fixed intervals of time by using Ar+ ion sputtering. The XPS spectra recorded after each etching step are placed in individual panel of Fig. 4 from bottom to up direction with upward shift for clarity. Besides the oxide peaks at the surface, the formation of TiN is confirmed by both Ti 2p and N 1s spectra for all the four samples. In Fig. 4(a), the binding energy related to Ti $2p_{3/2}$ in TiN varies between 454.75 eV to 454.60 eV and the Ti $2p_{3/2}$ in $TiSi_2$ comes at ~ 454 eV [43, 47] as shown by the violet dashed vertical line close to TiN peak. As the binding energy positions for TiN and $TiSi_2$ appear closely in Ti 2p spectra, the binding energy spectra of Si 2p can confirm the presence of the silicide phase. Interestingly, the dominance of silicide ($TiSi_2$) peak for all the four samples is evident in Si 2p spectra which is in contrary to the appearance of elemental silicon peak for samples annealed at 800°C and above (Fig. 3 and Fig. S1 in Supplementary Material). As we move towards the interface by etching out the film, the relative intensity (intensity of a peak with respect to its background intensity) of peaks related to TiN in Ti 2p and N 1s spectra decreases and the substrate peaks related to $Si_3N_4$ start to show up in both N 1s and Si 2p spectra at ~ 398.2 eV and ~ 102.3 eV, respectively. However, for silicide phase in the Si 2p spectra, the relative intensity initially increases as we etch out and move into the film and after attaining a maximum value, it starts to decrease and eventually the substrate peak takes over at the interface. Unlike for the samples presented in Fig 3, here, across the interface, no peak shifting towards higher energy is observed except for the sample annealed at 700°C. For this sample, an extra peak appears distinctly at higher binding energy at 402.0 eV in N 1s and at 106.6 eV in Si 2p spectra, as presented in Figs. 4 (h)&(i), respectively. Molecular nitrogen trapped at the interface can contribute to 402.0 eV [54, 63] while , the broad peak at about106.6 eV in Si 2p spectra is most likely related to amorphous silicon oxide [64] Further, these peaks might relate to the formation of Si-N-O based oxynitride phases too [54, 55]. However contrary to Fig. 3, here the peaks related to $Si_3N_4$ remain much stronger than that of oxynitride phases. Finally, with the help of a detailed XPS analysis presented in Figs. 3 & 4, we confirm the formation of TiN as the majority phase. With $T_a \geq 800°C$, decomposed silicon from $Si_3N_4$ substrate prefers to stay in its elemental form and a negligible amount may take part into the silicide formation, whereas, for $T_a < 800°C$, silicide ($TiSi_2$) formation is evident along with TiN. The quantitative analysis of nitride and silicide phases will be followed up later in the article.

The atomic percentage profile of the elements Ti, N, Si, O and C present in the studied samples are presented in Fig. 5 which exhibits the data from six samples representing six different annealing temperatures, namely, 820°C, 800°C, 780°C, 750°C, 700°C and 650°C, in (a), (b), (c), (d), (e) and (f), respectively. The core level binding energy spectra of Ti 2p, N 1s, Si 2p and O1s for these samples have been displayed in Figs. 3 & 4 and in Figs. S1 & S2 in Supplementary Material . The growth conditions, thickness of Ti films and annealing conditions except for $T_a$ were maintained at the closest possible level among the samples. Here, the elemental atomic (%) is presented with respect to etching time which relates the removal of films in a controlled manner by using Ar+ ion sputtering. Dominance of oxygen level at the initial stages, as evident in the atomic profile of all the samples, can be mainly due to the air exposure of the samples before loading into the XPS chamber and it is supported by the appearance of oxide peaks on the surface of the samples as presented in the previous two figures. However, introduction of oxygen contamination to the film is inevitable during the growth and annealing process that were carried out in the vacuum environment with base pressure less than $1.5 \times 10^{-7}$ Torr. Further, the presence of trapped oxygen molecule at the interface of $Si_3N_4$/Si might also contribute to the oxygen content in the film. The formation of oxynitride and oxide phases at the interface also suggests the presence of oxygen inside the film and at the interface.

For all the samples, with progressing the etching time up to a certain point, atomic percentage of Ti shows very little variation indicating the ejected photoelectrons carry the information about the films only. Accordingly, we have marked this extent by black vertical lines to indicate the boundary of the film and the interface in Fig.5. During this time range we observe the presence and variation of N and Si atoms that eventually diffused from the substrate into Ti film after getting thermally decomposed from $Si_3N_4$ layer of the substrate. With further etching, percentage of Ti starts to decrease rapidly while Si and N percentages show up an increasing trend. This indicates that the XPS spectra start to collect information from $Si_3N_4$ substrate in addition to the transformed composite film and hence the film-substrate interface properties are embedded into the measured XPS data. The two dashed vertical lines at 150 sec and 360 sec refer to the contributions from the film and the interface, respectively, and a more detailed analysis at these two pints are discussed in the next section of the article. It is very clear from Fig. 5 that the atomic percentage of N and Si and their variation across the film depend strongly on the annealing temperature. For $T_a \geq 800°C$ as shown in Figs. 5(a)&(b), atomic percentage of N across the film side is all the time more than that of Si, and this causes the nitride phase as the majority phase as appeared in the binding energy spectra presented in Fig. 3 and also in the XRD pattern in Fig. 2. However, as the annealing temperature decreases the gap between the atomic percentage of N and Si gets reduced and at $T_a \leq 780°C$, in the middle of the film Si percentage takes over and stays higher than that of N for an extent before the latter becomes greater. Further, with lowering $T_a$, overall percentage of Si and N gets lowered and at around 650°C, both become reasonably less and almost merge with each other as observed in Fig. 5(f). The silicide formation at $T_a \leq$ 780°C, as evident in Fig. 4 (third column: Si 2p spectra), can be understood with the atomic profile presented here in Fig. 5. Here, the availability of Si is more or comparable with that of N. In order to form the stable $TiSi_2$ phase, one-part Ti requires two-parts of Si, whereas, only one atom from each N and Ti can form easily stoichiometric TiN. Therefore, the displayed atomic profile explains clearly the annealing temperature dependent decomposition of $Si_3N_4$ in presence of Ti and diffusion mediated relative variation in decomposed Si and N atoms across the Ti thin film and at the interface.

The comparison of elemental atomic percentage for all the elements present in individual sample for a particular $T_a$ has been displayed in Fig. 5. However, this is important to know how the atomic concentration for individual element varies and evolves with $T_a$ as that directly relates to the availability of a particular element to take part into a chemical reaction with other elements present in a sample. Hence, as discussed in the present study, the formation processes of nitride, silicide, oxynitride and oxide phases strongly depend on the concentration of the constituent elements across the films. Hence for a better comparison, we have gathered the atomic percentage variations of *an individual element* for all the measured samples together. Accordingly, in Fig. 6, we compare the variation of atomic concentration across the films for three most influencing elements, viz. O, N and Si, separately in (a), (b) and (c), respectively. For the variation of oxygen concentration level with etch time as presented in Fig. 6(a), an overall increasing trend for oxygen concentration is observed with decreasing annealing temperature. With $T_a$ = 800°C & 820°C, we observe no noticeable change in the oxygen concentration as they merge on each other. For $T_a \leq$ 780°C, the oxygen concentration on the sample surface (for the as-loaded sample before etching started) raises to a maximum value of ~ 45 at. % and the concentration remains significant for wider span of etching time with decreasing annealing temperature. This indicates that the probability of oxide and oxynitride formation, on sample surface as well as inside the film, increases for reduced $T_a$. Whereas, with increasing $T_a$, nitrogen concentration increases across the film before reaching the metal-substrate interface as shown in Fig. 6(b). Intermediate shallow type of structure can be related to increased Si concentration as observed in Fig. 6(c). In contrary to the trend followed by oxygen and nitrogen, silicon concentration varies in two ways with $T_a$; first it increases with decreasing $T_a$ in the range from 820°C-to- 800°C-to-780°C, and then it decreases with decreasing $T_a$ from 750°C-to- 700°C-to-650°C. The maximum Si concentration inside the film occurs for $T_a$ = 780°C. However, the change in Si concentration due to annealing temperature is less strong than that observed for nitrogen concentration. Further, We have compared the core level binding energy spectra for Ti 2p, N 1s and Si 2p among the samples in (d), (e) and (f) of Fig. 6, respectively. Here we have selected the spectra obtained after 150 sec of Ar+ ion etching, as at this point the oxygen concentration is reasonably less for all the samples and also the influence of substrate can be ignored. A little shift in binding energy related to TiN is observed in Ti 2p spectra where the shifting is towards lower binding energy for $T_a$ decreasing from 820°C-to-800°C-to-780°C, and then it shifts towards higher binding energy with decreasing $T_a$ from 750°C-to-700°C-to-650°C. However, N 1s spectra shown in Fig. 6(e) does not show any shift and only the nitride phase is evident. Whereas, as we move towards lower $T_a$ from 800°C to 780°C, Si 2p spectra presented in Fig. 6(f) show a distinct shift in binding energy corresponding to the dominance of silicide phase over that of elemental silicon. Therefore, it is evident that annealing temperature plays a crucial role, particularly, in the formation of silicide phase. Further, the little shift in Ti 2p with the annealing temperature can be due to the presence of silicide phase in greater way than that for $T_a \geq$ 800°C.

The presence of different phases formed during the annealing process and their quantitative analysis have been carried out by performing the spectra deconvolution using a mixed Gaussian/Lorentzian based peak fitting model in the framework of Avantage Software with Shirley background correction. The fitted graphs for Ti 2p, N 1s and Si 2p are shown in Fig. 7 which shows the data for the measured six samples represented by their respective annealing temperature as marked in the figure. Here, the displayed data corresponds to the XPS measurements obtained after etching the samples for 150 sec, as at this point, the information can be mostly from the film with almost no contribution from the substrate. The position related to 150 sec etching time is shown by the dashed vertical line in the atomic profile displayed in Fig. 5. The presentation in Fig. 7 is organised as, each row representing the spectra for the afore-mentioned three elements for a particular sample and each column displaying spectra for a specific element as obtained from all the measured samples. For Ti 2p, three main phases are considered as they are supported by the other elements in their respective XPS spectra and the fitting is carried out in accordance with the common procedure followed in the literature [65]. The phases are nitride (TiN), silicide ($TiSi_2$) and a collective oxygenated phase (Ti-N-O) [17, 62, 66] combining the possible oxynitride ($TiN_xO_y$) and oxide ($TiO_2$) phases due to the presence of oxygen during growth, annealing and exposure to air. TiN and $TiSi_2$ phases are justified by N 1s and Si 2p spectra as shown in the middle column and the right column, respectively. Apart from TiN (~ 397 eV) [67] and Ti-O-N (~ 397.5 eV) [14, 47, 61, 68] peaks in N1s, a low intensity broad peak appearing at higher binding energy (>399 eV) for N 1s can be related to oxynitride (N-O) bonding [40, 63, 69]. This N-O bonding may originate from Si oxynitride (Si-O-N) [55, 70] and/or may be from titanium oxynitride (Ti-O-N) [63, 71]. Similarly, for Si 2p as appeared in many of the samples, the peak at ~ 102 eV might correspond to the formation of silicon oxynitride (Si-N-O) [54] and/or oxide phase ($SiO_x$) [47]. Here, it should be noted that 102 eV can be related to $Si_3N_4$ also but as we are at the middle of the film, we can ignore the effect of the substrate $Si_3N_4$ layer [51]. The fitting parameters and the atomic percentage of the phases are listed in Table-S1 shown in Supplementary Material. It is evident from fitting that the nitride phase for all the samples represents the majority phase compared to that of silicide phase. Furthermore, we have compared our binding energy values, obtained from the deconvolution of the spectra as presented in Fig. 7, with the reported values from the literature and the comparison is displayed in tabular form in Table 1. It is evident that the obtained values are in accordance with the literature reported values and hence the peak assignments are justified.

Finally, we have summarized the key observations by a quantitative analysis of the nitride and silicide phases formed in different samples. The relative concentration (%) of the two aforementioned phases with respect to etch time for all the six samples are shown in Fig. 8. The analysis is done based on the fitting as illustrated in Fig. 7. Here, we have obtained the relative concentration from the fitted area of the deconvoluted spectra of Ti 2p for TiN and $TiSi_2$ phases up to etching time of 360 sec at which point $Si_3N_4$ substrate peaks start to appear in the XPS spectra of Si 2p and N 1s. Similar to 150 sec etching time, 360 sec is also marked in the atomic profile presented in Fig. 5 by dashed vertical line. In Fig. 8, the red closed circular points indicate the percentage of TiN phase while, the blue open circles represent the amount of $TiSi_2$ phase present in the sample. First of all, TiN phase dominates over the $TiSi_2$ phase for all the studied annealing temperature across the films. Further, we observe that the overall gap or the difference between the percentage of nitride and silicide phases decrease with annealing temperature in the temperature range 820°C to 780°C and at 780°C, the silicide phase reaches at maximum among the samples and the difference is the minimum. We have already observed in Figs. 5&6 that amount of silicon appears to be the maximum for the sample annealed at 780°C among all the measured samples, hence it is expected that the silicide formation would be maximum for $T_a$ = 780°C. For 650°C ≤ $T_a$ ≤ 780°C, the gap between the silicide and nitride phase mostly follows in reverse way, i.e. the difference increases with reduced $T_a$. Here, overall availability of decomposed nitrogen and silicon atoms gets reduced with lower $T_a$ and hence probability of forming $TiSi_2$ with two Si atoms worsened compared to the probability of forming TiN with one nitrogen atom. Further from Fig. 6, we have observed that the binding energy corresponding to Ti 2p in TiN shifted towards lower binding energy as we reduce the $T_a$ from 820°C to 780°C and the shift is maximum for $T_a$ = 780°C and then the binding energy shifts towards higher energy again for further lowering the $T_a$ from 780°C to 650°C. This shifting of binding energy for Ti 2p in the nitride phase with $T_a$ resembles the variation in the gap between the nitride and silicide phases with $T_a$. Hence the binding energy shifting might be originated or caused by the amount of silicide phase present in a particular sample. The presence of silicide phase is most prominent at $T_a$ = 780°C and the binding energy shift is also the strongest for this case. Besides, binding energy shifting towards lower energy might indicate the formation of non-stoichiometric nitride following the availability of reduced number of nitrogen atoms at lower $T_a$. However, binding energy shifting due to the non-stoichiometry would then follow a similar trend with the change in annealing temperature, i.e., shifting towards lower binding energy would be continuously increasing with decreasing $T_a$. Further, the relative concentration of $TiSi_2$ for lower $T_a$ can include some contribution from metallic Ti also as the presence of metallic Ti is evident from the surface scan. The binding energy difference in metallic Ti and $TiSi_2$ is about 0.3 eV [43] and therefore, overlapping is expected for these two phases. For simplicity, we have only considered $TiSi_2$ phase and ignored the contribution from the metallic Ti during the deconvolution process.

## 4. Conclusions

In conclusion, we have presented Ti/$Si_3N_4$ as a model metal-substrate assembly to probe the interaction between metal and the substrate during high temperature annealing under high vacuum environment. Decomposition of $Si_3N_4$ into Si and N atoms followed by interdiffusion under thermal treatment is the key mechanism behind the metal-substrate interaction. The reactive nature of Ti towards N and Si has been established by using XPS based interface and depth profile studies. By reacting with N and Si, Ti forms its stable binary compounds TiN and $TiSi_2$ respectively. A quantitative analysis of these two phases along with their dependence on the annealing temperature has been carried out for the first time. Here, we have demonstrated that the amount of the afore-mentioned phases strongly depends on the annealing temperature. However, each of the studied cases unambiguously reflects TiN as the majority phase. Interestingly, we observe that at $T_a \geq 800$°C, majority of the decomposed Si atoms prefer to stay in its elemental form with a very little contribution towards the silicide formation. The diffusion occurs maximum for decomposed Si into Ti film at $T_a$= 780°C which also leads to the maximum amount of silicide formation. Further, the amount of silicide phase is observed to influence the binding energy of Ti 2p in TiN, which gets shifted towards lower energy with increasing amount of silicide phase.

Moreover, as TiN appears to be the majority phase, the presented substrate mediated nitridation technique can be adopted to fabricate stoichiometric TiN which is known for its versatile functionalities in the field of superconductivity [72, 73], optics [74], plasmonics [75-78], ceramics [79, 80] and in many other multidisciplinary areas [81-85]. Besides, the combination of TiN and $TiSi_2$ based binary composites can eventually serve as Ti-Si-N based ternary family of composite materials that can be produced without the presence of any external source of N and Si. Finally, one of our main interests coincides with the fabrication of disordered superconducting TiN thin films using this technique which inherently produces disorder by incorporating non-superconducting component like Si and the silicide phase. Eventually, tuning the amount of $TiSi_2$ by varying the annealing temperature one can introduce disorder, scattering centres, defects etc. into TiN to explore many of the excellent and unique properties possessed by the latter in an advantageous way.

**Acknowledgements**

The authors acknowledge the technical support in carrying out the XPS measurements using the central facility at IIT, Mandi, India. Mr. Naveen Gumra from IIT Mandi is highly acknowledged for his assistance in XPS characterization. We are thankful to Dr. K.K. Maurya for providing technical help in characterizing thin films through HRXRD. We are thankful to Mr. M. B. Chhetri, Mr. Bikash Gajar and Ms. Deepika Sawle for their assistance in the lab. S.Y. acknowledges the Senior Research Fellowship (SRF) from UGC. This work was supported by CSIR network project 'AQuaRIUS' (Project No. PSC 0110) and is carried out under the mission mode project "Quantum Current Metrology".

**References:**

[1] S.-L. Zhang, M. Östling, Metal Silicides in CMOS Technology: Past, Present, and Future Trends, Crit. Rev. Solid State 28 (2003) 1-129.

[2] C. Ren, B.B. Faizhal, D.S.H. Chan, M.F. Li, Y.C. Yeo, A.D. Trigg, N. Balasubramanian, D.L. Kwong, Work function tuning of metal nitride electrodes for advanced CMOS devices, Thin Solid Films 504 (2006) 174-177.

[3] A. Kikuchi, T. Ishiba, Role of oxygen and nitrogen in the titanium-silicon reaction, J. Appl. Phys. 61 (1987) 1891-1894.

[4] A.E. Morgan, E.K. Broadbent, D.K. Sadana, Reaction of titanium with silicon nitride under rapid thermal annealing, Appl. Phys. Lett. 49 (1986) 1236-1238.

[5] F.M. d'Heurle, P. Gas, Kinetics of formation of silicides: J. Mater. Res. 1 (1986) 205-221.

[6] A.E. Morgan, E.K. Broadbent, K.N. Ritz, D.K. Sadana, B.J. Burrow, Interactions of thin Ti films with Si, $SiO_2$, $Si_3N_4$, and $SiO_xN_y$ under rapid thermal annealing, J. Appl. Phys. 64 (1988) 344-353.

[7] G. Gagnon, J.F. Currie, J.L. Brebner, T. Darwall, Efficiency of TiN diffusion barrier between Al and Si prepared by reactive evaporation and rapid thermal annealing, J. Appl. Phys. 79 (1996) 7612-7620.

[8] M.M. Frank, C. Cabral, J.M. Dechene, C. Ortolland, Y. Zhu, E.D. Marshall, C.E. Murray, M.P. Chudzik, Titanium Silicide/Titanium Nitride Full Metal Gates for Dual-Channel Gate-First CMOS, IEEE Electron Device Lett. 37 (2016) 150-153.

[9] T. Hara, K. Tani, K. Inoue, S. Nakamura, T. Murai, Formation of titanium nitride layers by the nitridation of titanium in high-pressure ammonium ambient, Appl. Phys. Lett. 57 (1990) 1660-1662.

[10] J.C. Barbour, A.E.T. Kuiper, M.F.C. Willemsen, A.H. Reader, Thin-film reaction between Ti and $Si_3N_4$, Appl. Phys. Lett. 50 (1987) 953-955.

[11] S. Yadav, A. Sharma, B. Gajar, M. Kaur, D. Singh, S. Singh, K.K. Maurya, S. Husale, V.N. Ojha, S. Sahoo, Substrate Mediated Synthesis of Ti–Si–N Nano-and-Micro Structures for Optoelectronic Applications, Adv. Eng. Mater. 21 (2019) 1900061.

[12] N.A. Saveskul, N.A. Titova, E.M. Baeva, A.V. Semenov, A.V. Lubenchenko, S. Saha, H. Reddy, S.I. Bogdanov, E.E. Marinero, V.M. Shalaev, A. Boltasseva, V.S. Khrapai, A.I. Kardakova, G.N. Goltsman, Superconductivity Behavior in Epitaxial TiN Films Points to Surface Magnetic Disorder, Phys. Rev. Appl. 12 (2019) 054001.

[13] T.I. Baturina, S.V. Postolova, A.Y. Mironov, A. Glatz, M.R. Baklanov, V.M. Vinokur, Superconducting phase transitions in ultrathin TiN films, Europhys. Lett. 97 (2012) 17012.

[14] D. Shah, A. Catellani, H. Reddy, N. Kinsey, V. Shalaev, A. Boltasseva, A. Calzolari, Controlling the Plasmonic Properties of Ultrathin TiN Films at the Atomic Level, ACS Photonics 5 (2018) 2816-2824.

[15] C.-C. Chang, J. Nogan, Z.-P. Yang, W.J.M. Kort-Kamp, W. Ross, T.S. Luk, D.A.R. Dalvit, A.K. Azad, H.-T. Chen, Highly Plasmonic Titanium Nitride by Room-Temperature Sputtering, Sci. Rep. 9 (2019) 15287.

[16] G.V. Naik, V.M. Shalaev, A. Boltasseva, Alternative Plasmonic Materials: Beyond Gold and Silver, Adv. Mater. 25 (2013) 3264-3294.

[17] L. Duta, E.G. Stan, C.A. Popa, A.M. Husanu, S. Moga, M. Socol, I. Zgura, F. Miculescu, I. Urzica, C.A. Popescu, N.I. Mihailescu, Thickness Influence on In Vitro Biocompatibility of Titanium Nitride Thin Films Synthesized by Pulsed Laser Deposition, Materials 9 (2016) 38.

[18] J. Fukushima, K. Kashimura, H. Takizawa, Nitridation Reaction of Titanium Powders by 2.45 GHz Multimode Microwave Irradiation using a SiC Susceptor in Atmospheric Conditions, Processes 8 (2020) 20.

[19] S. Oktay, Z. Kahraman, M. Urgen, K. Kazmanli, XPS investigations of tribolayers formed on TiN and (Ti,Re)N coatings, Appl. Surf. Sci. 328 (2015) 255-261.

[20] B. Gajar, S. Yadav, D. Sawle, K.K. Maurya, A. Gupta, R.P. Aloysius, S. Sahoo, Substrate mediated nitridation of niobium into superconducting $Nb_2N$ thin films for phase slip study, Sci. Rep. 9 (2019) 8811.

[21] H.D. Batha, E.D. Whitney, Kinetics and Mechanism of the Thermal Decomposition of $Si_3N_4$, J. Am. Ceram. Soc. 56 (1973) 365-369.

[22] I. Jauberteau, R. Mayet, J. Cornette, D. Mangin, A. Bessaudou, P. Carles, J.L. Jauberteau, A. Passelergue, Silicides and Nitrides Formation in Ti Films Coated on Si and Exposed to ($Ar$-$N_2$-$H_2$) Expanding Plasma, Coatings 7 (2017) 23.

[23] A. Torgovkin, S. Chaudhuri, A. Ruhtinas, M. Lahtinen, T. Sajavaara, I.J. Maasilta, High quality superconducting titanium nitride thin film growth using infrared pulsed laser deposition, Supercond. Sci. Technol. 31 (2018) 055017.

[24] A. Kardakova, M. Finkel, D. Morozov, V. Kovalyuk, P. An, C. Dunscombe, M. Tarkhov, P. Mauskopf, T.M. Klapwijk, G. Goltsman, The electron-phonon relaxation time in thin superconducting titanium nitride films, Appl. Phys. Lett. 103 (2013) 252602.

[25] J.B. Chang, M.R. Vissers, A.D. Córcoles, M. Sandberg, J. Gao, D.W. Abraham, J.M. Chow, J.M. Gambetta, M. Beth Rothwell, G.A. Keefe, M. Steffen, D.P. Pappas, Improved superconducting qubit coherence using titanium nitride, Appl. Phys. Lett. 103 (2013) 012602.

[26] A. Shearrow, G. Koolstra, S.J. Whiteley, N. Earnest, P.S. Barry, F.J. Heremans, D.D. Awschalom, E. Shirokoff, D.I. Schuster, Atomic layer deposition of titanium nitride for quantum circuits, Appl. Phys. Lett. 113 (2018) 212601.

[27] B. Sacépé, C. Chapelier, T.I. Baturina, V.M. Vinokur, M.R. Baklanov, M. Sanquer, Disorder-Induced Inhomogeneities of the Superconducting State Close to the Superconductor-Insulator Transition, Phys. Rev. Lett. 101 (2008) 157006.

[28] T.I. Baturina, A.Y. Mironov, V.M. Vinokur, M.R. Baklanov, C. Strunk, Localized Superconductivity in the Quantum-Critical Region of the Disorder-Driven Superconductor-Insulator Transition in TiN Thin Films, Phys. Rev. Lett. 99 (2007) 257003.

[29] I. Schneider, K. Kronfeldner, T.I. Baturina, C. Strunk, Quantum phase slips and number-phase duality in disordered TiN nanostrips, Phys. Rev. B 99 (2019) 094522.

[30] R. Simpson, R.G. White, J.F. Watts, M.A. Baker, XPS investigation of monatomic and cluster argon ion sputtering of tantalum pentoxide, Appl. Surf. Sci. 405 (2017) 79-87.

[31] M. Maeda, R. Oomoto, T. Shibayanagi, M. Naka, Solid-state diffusion bonding of silicon nitride using titanium foils, Metall. Mater. Trans. A 34 (2003) 1647-1656.

[32] M. Paulasto, J.K. Kivilahti, F.J.J. van Loo, Interfacial reactions in Ti/Si3N4 and TiN/Si diffusion couples, J. Appl. Phys. 77 (1995) 4412-4416.

[33] Malcolm W. Chase, Jr., NIST-JANAF thermochemical tables, Fourth edition. Washington, DC : American Chemical Society ; New York : American Institute of Physics for the National Institute of Standards and Technology, 1998.1998.

[34] I. Barin, O. Knacke, O. Kubaschewski, Thermochemical properties of inorganic substances, 1st ed., Springer-Verlag Berlin Heidelberg1977.

[35] H.M. Jennings, On reactions between silicon and nitrogen, J. Mater. Sci. 18 (1983) 951-967.

[36] S.C. Singhal, Thermodynamic analysis of the high-temperature stability of silicon nitride and silicon carbide, Ceramurgia International 2 (1976) 123-130.

[37] X. Hu, C. Shao, J. Wang, H. Wang, Characterization and high-temperature degradation mechanism of continuous silicon nitride fibers, J. Mater. Sci. 52 (2017) 7555-7566.

[38] A.S. Bhansali, R. Sinclair, A.E. Morgan, A thermodynamic approach for interpreting metallization layer stability and thin-film reactions involving four elements: Application to integrated circuit contact metallurgy, J. Appl. Phys. 68 (1990) 1043-1049.

[39] Y.C. Ee, Z. Chen, L. Chan, K.H. See, S.B. Law, S. Xu, Z.L. Tsakadze, P.P. Rutkevych, K.Y. Zeng, L. Shen, Formation of Ti–Si–N film using low frequency, high density inductively coupled plasma process, Journal of Vacuum Science & Technology B: Microelectronics and Nanometer Structures Processing, Measurement, and Phenomena 23 (2005) 2444-2448.

[40] D. Jaeger, J. Patscheider, A complete and self-consistent evaluation of XPS spectra of TiN, J. Electron Spectrosc. 185 (2012) 523-534.

[41] J.T. Mayer, U. Diebold, T.E. Madey, E. Garfunkel, Titanium and reduced titania overlayers on titanium dioxide(110), J. Electron Spectrosc. 73 (1995) 1-11.

[42] R. Gouttebaron, D. Cornelissen, R. Snyders, J.P. Dauchot, M. Wautelet, M. Hecq, XPS study of TiOx thin films prepared by d.c. magnetron sputtering in Ar–$O_2$ gas mixtures, Surf. Interface Anal. 30 (2000) 527-530.

[43] P.L. Tam, Y. Cao, L. Nyborg, XRD and XPS characterisation of transition metal silicide thin films, Surf. Sci. 606 (2012) 329-336.

[44] K.S. Robinson, P.M.A. Sherwood, X-Ray photoelectron spectroscopic studies of the surface of sputter ion plated films, Surf. Interface Anal. 6 (1984) 261-266.

[45] G. Greczynski, D. Primetzhofer, J. Lu, L. Hultman, Core-level spectra and binding energies of transition metal nitrides by non-destructive x-ray photoelectron spectroscopy through capping layers, Appl. Surf. Sci. 396 (2017) 347-358.

[46] P. Prieto, R.E. Kirby, X-ray photoelectron spectroscopy study of the difference between reactively evaporated and direct sputter-deposited TiN films and their oxidation properties, J. Vac. Sci. Technol. A 13 (1995) 2819-2826.

[47] NIST X-ray Photoelectron Spectroscopy Database, Version 4.1 (National Institute of Standards and Technology, Gaithersburg, 2012); http://srdata.nist.gov/xps/.

[48] H.C. Barshilia, M. Ghosh, Shashidhara, R. Ramakrishna, K.S. Rajam, Deposition and characterization of TiAlSiN nanocomposite coatings prepared by reactive pulsed direct current unbalanced magnetron sputtering, Appl. Surf. Sci. 256 (2010) 6420-6426.

[49] S.M. Lee, E.T. Ada, H. Lee, J. Kulik, J.W. Rabalais, Growth of Ti and $TiSi_2$ films on Si(111) by low energy Ti+ beam deposition, Surf. Sci. 453 (2000) 159-170.

[50] M.C. Poon, C.W. Kok, H. Wong, P.J. Chan, Bonding structures of silicon oxynitride prepared by oxidation of Si-rich silicon nitride, Thin Solid Films 462-463 (2004) 42-45.

[51] F.N. Cubaynes, V.C. Venezia, C. van der Marel, J.H.M. Snijders, J.L. Everaert, X. Shi, A. Rothschild, M. Schaekers, Plasma-nitrided silicon-rich oxide as an extension to ultrathin nitrided oxide gate dielectrics, Appl. Phys. Lett. 86 (2005) 172903.

[52] F.J. Himpsel, F.R. McFeely, A. Taleb-Ibrahimi, J.A. Yarmoff, G. Hollinger, Microscopic structure of the $SiO_2$/Si interface, Phys. Rev. B 38 (1988) 6084-6096.
[53] R. Klauser, I.H. Hong, H.J. Su, T.T. Chen, S. Gwo, S.C. Wang, T.J. Chuang, V.A. Gritsenko, Oxidation states in scanning-probe-induced $Si_3N_4$ to $SiO_x$ conversion studied by scanning photoemission microscopy, Appl. Phys. Lett. 79 (2001) 3143-3145.
[54] X. Zhang, S. Ptasinska, Growth of silicon oxynitride films by atmospheric pressure plasma jet, J. Phys. D: Applied Physics 47 (2014) 145202.
[55] J. Eng, I.A. Hubner, J. Barriocanal, R.L. Opila, D.J. Doren, X-ray photoelectron spectroscopy of nitromethane adsorption products on Si(100): A model for N1s core-level shifts in silicon oxynitride films, J. Appl. Phys. 95 (2004) 1963-1968.
[56] Chien-Jen Tang, Cheng-Chung Jaing, Chuen-Lin Tien, Wei-Chiang sun, Shih-Chin Lin, Optical, structural, and mechanical properties of silicon oxynitride films prepared by pulsed magnetron sputtering, Appl. Opt. 56 (2017) C168.
[57] A. Akkaya, B. Boyarbay, H. Çetin, K. Yıldızlı, E. Ayyıldız, A Study on the Electronic Properties of $SiO_xN_y$/p-Si Interface, Silicon 10 (2018) 2717-2725.
[58] W.Y. Yang, H. Iwakuro, H. Yagi, T. Kuroda, S. Nakamura, Study of Oxidation of TiSi2Thin Film by XPS, Japanese J. Appl. Phys. 23 (1984) 1560-1567.
[59] A.A. Galuska, J.C. Uht, N. Marquez, Reactive and nonreactive ion mixing of Ti films on carbon substrates, J. Vac. Sci. Technol. A 6 (1988) 110-122.
[60] R. Huang, Z. Lin, Y. Guo, C. Song, X. Wang, H. Lin, L. Xu, J. Song, H. Li, Bright red, orange-yellow and white switching photoluminescence from silicon oxynitride films with fast decay dynamics, Opt. Mater. Express 4 (2014) 205-212.
[61] M.V. Kuznetsov, J.F. Zhuravlev, V.A. Gubanov, XPS analysis of adsorption of oxygen molecules on the surface of Ti and $TiN_x$ films in vacuum, J. Electron Spectrosc. 58 (1992) 169-176.

[62] I. Bertóti, M. Mohai, J.L. Sullivan, S.O. Saied, Surface characterisation of plasma-nitrided titanium: an XPS study, Appl. Surf. Sci. 84 (1995) 357-371.

[63] I. Milošv, H.H. Strehblow, B. Navinšek, M. Metikoš-Huković, Electrochemical and thermal oxidation of TiN coatings studied by XPS, Surf. Interface Anal. 23 (1995) 529-539.

[64] A.U. Alam, M.M.R. Howlader, M.J. Deen, Oxygen Plasma and Humidity Dependent Surface Analysis of Silicon, Silicon Dioxide and Glass for Direct Wafer Bonding, ECS J. Solid State Sci. Technol. 2 (2013) P515-P523.

[65] G. Greczynski, L. Hultman, Self-consistent modelling of X-ray photoelectron spectra from air-exposed polycrystalline TiN thin films, Appl. Surf. Sci. 387 (2016) 294-300.

[66] R.R. Mohanta, V.R.R. Medicherla, K.L. Mohanta, N.C. Nayak, S. Majumder, V. Solanki, S. Varma, K. Bapna, D.M. Phase, V. Sathe, Ion beam induced chemical and morphological changes in $TiO_2$ films deposited on Si(111) surface by pulsed laser deposition, Appl. Surf. Sci. 325 (2015) 185-191.

[67] S. Hammouti, B. Holybee, W. Zhu, J.P. Allain, B. Jurczyk, D.N. Ruzic, Titanium nitride formation by a dual-stage femtosecond laser process, Applied Physics A 124 (2018) 411.

[68] M.V. Kuznetsov, J.F. Zhuravlev, V.A. Zhilyaev, V.A. Gubanov, XPS study of the nitrides, oxides and oxynitrides of titanium, J. Electron Spectrosc 58 (1992) 1-9.

[69] J.C. Oliveira, F. Fernandes, F. Ferreira, A. Cavaleiro, Tailoring the nanostructure of Ti–Si–N thin films by HiPIMS in deep oscillation magnetron sputtering (DOMS) mode, Surf. Coat. Technol. 264 (2015) 140-149.

[70] A.R. Chourasia, D.R. Chopra, X-ray photoelectron study of TiN/$SiO_2$ and TiN/Si interfaces, Thin Solid Films 266 (1995) 298-301.

[71] J. Halbritter, H. Leiste, H.J. Mathes, P. Walk, ARXPS — Studies of nucleation and make-up of sputtered TiN-layers, Fresenius' J. Anal. Chem. 341 (1991) 320-324.

[72] P.C.J.J. Coumou, E.F.C. Driessen, J. Bueno, C. Chapelier, T.M. Klapwijk, Electrodynamic response and local tunneling spectroscopy of strongly disordered superconducting TiN films, Phys. Rev. B 88 (2013) 180505.

[73] S.E. de Graaf, R. Shaikhaidarov, T. Lindström, A.Y. Tzalenchuk, O.V. Astafiev, Charge control of blockade of Cooper pair tunneling in highly disordered TiN nanowires in an inductive environment, Phys. Rev. B 99 (2019) 205115.

[74] L. Peng, X. Wang, I. Coropceanu, A.B. Martinson, H. Wang, D.V. Talapin, X. Ma, Titanium Nitride Modified Photoluminescence from Single Semiconductor Nanoplatelets, Adv. Funct. Mater. 30 (2020) 1904179.

[75] G.V. Naik, B. Saha, J. Liu, S.M. Saber, E.A. Stach, J.M.K. Irudayaraj, T.D. Sands, V.M. Shalaev, A. Boltasseva, Epitaxial superlattices with titanium nitride as a plasmonic component for optical hyperbolic metamaterials, Pro. Natl. Acad. Sci. 111 (2014) 7546.

[76] S. Ishii, R.P. Sugavaneshwar, T. Nagao, Titanium Nitride Nanoparticles as Plasmonic Solar Heat Transducers, J. Phys. Chem. C 120 (2016) 2343-2348.

[77] L. Gui, S. Bagheri, N. Strohfeldt, M. Hentschel, C.M. Zgrabik, B. Metzger, H. Linnenbank, E.L. Hu, H. Giessen, Nonlinear Refractory Plasmonics with Titanium Nitride Nanoantennas, Nano Lett. 16 (2016) 5708-5713.

[78] A.A. Barragan, S. Hanukovich, K. Bozhilov, S.S.R.K.C. Yamijala, B.M. Wong, P. Christopher, L. Mangolini, Photochemistry of Plasmonic Titanium Nitride Nanocrystals, J. Phys. Chem. C 123 (2019) 21796-21804.

[79] I.A. Kovalev, A.I. Ogarkov, A.V. Shokod'ko, S.V. Shevtsov, A.A. Konovalov, S.V. Kannykin, A.A. Ashmarin, G.P. Kochanov, A.S. Chernyavskii, K.A. Solntsev, Structural and Phase Transformations in Compact Titanium Nitride-Based Ceramics during High-Temperature Heating in Gaseous Media, Inorg. Mater. 55 (2019) 851-855.

[80] A.S. Chernyavskii, Synthesis of Ceramics Based on Titanium, Zirconium, and Hafnium Nitrides, Inorg. Mater. 55 (2019) 1303-1327.

[81] R.P. van Hove, I.N. Sierevelt, B.J. van Royen, P.A. Nolte, Titanium-Nitride Coating of Orthopaedic Implants: A Review of the Literature, Biomed Res. Int. 2015 (2015) 485975.

[82] G. Qiu, A. Thakur, C. Xu, S.-P. Ng, Y. Lee, C.-M.L. Wu, Detection of Glioma-Derived Exosomes with the Biotinylated Antibody-Functionalized Titanium Nitride Plasmonic Biosensor, Adv. Funct. Mater. 29 (2019) 1806761.

[83] W.-G. Lim, C. Jo, A. Cho, J. Hwang, S. Kim, J.W. Han, J. Lee, Approaching Ultrastable High-Rate Li–S Batteries through Hierarchically Porous Titanium Nitride Synthesized by Multiscale Phase Separation, Adv. Mater. 31 (2019) 1806547.

[84] J. Yu, P. Phang, C. Samundsett, R. Basnet, G.P. Neupan, X. Yang, D.H. Macdonald, Y. Wan, D. Yan, J. Ye, Titanium Nitride Electron-Conductive Contact for Silicon Solar Cells By Radio Frequency Sputtering from a TiN Target, ACS Appl. Mater. Interfaces 12 (2020) 26177-26183.

[85] J.M. Pauls, N.F. Shkodich, A.S. Mukasyan, Mechanisms of Self-Sustained Reaction in Mechanically Induced Nanocomposites: Titanium Nitride and Boron, J. Phys. Chem. C 123 (2019) 11273-11283.

**Figure Captions:**

**Fig. 1:** Schematic presentation of the metal-substrate interaction process under high temperature annealing. (a) Growth of Ti thin film on $Si_3N_4$/Si substrate by DC magnetron sputtering. (b) The annealing process with time dependent temperature monitoring is shown graphically with color-gradient lines. The middle rectangular reddish block represents the temperature range used in the present study for annealing. (c) Step-wise representation of the interaction between Ti and $Si_3N_4$ layers and subsequent interdiffusion of molecules during the annealing process. (d)Transformation of Ti thin films to a composite film of TiN and $TiSi_2$ by the annealing process explained in (b) and (c).

**Fig. 2:** XRD characterization of three representative samples annealed at temperatures ≥ 800 °C. Two of the three (two spectra from the bottom) samples were annealed at 820 °C and having thickness values as 25 nm and 15 nm, respectively. The top spectrum relates to a sample having thickness ~ 15 nm and annealed at 800 °C.

**Fig. 3:** X-ray photoelectron spectroscopy (XPS) analysis on a representative sample having thickness of 15 nm and annealed at 820°C for 2 hours. XPS characterization showing the core level binding energy of Ti 2p (a), N 1s (b), Si 2p (c) and O 1s (d). The XPS spectrum measured on the top surface of the as loaded sample is shown at the bottom and in the upward direction, the curves represent the scans measured after etching the film (in steps with constant time and energy) from the top surface till the substrate $Si_3N_4$/Si.

**Fig. 4:** XPS characterization of four representative reference samples annealed at 780°C (top row), 750°C (second row from the top), 700°C (third row from the top) and 650°C (bottom row), respectively. The samples are annealed for 2 hours individually and are of 15 nm thick each. Left column [(a), (d), (g) and (j)]: core level binding energy spectra of Ti 2p; Middle column [(b), (e), (h) and (k)]: spectra of N 1s; Right column [(c), (f), (i) and (l)]: the same for Si 2p. The XPS scan measured on the top surface of the as loaded sample is shown by the first spectrum from the bottom in each of the respective spectra presented in (a)-(l). For clarity the curves are shifted in the upward direction to represent the scans measured after

etching the film (in steps with constant time and energy) from the top surface until the substrate $Si_3N_4$/Si.

**Fig. 5:** Variation of atomic (%) profile with annealing temperature. Atomic percentage of Ti, N, Si, O and C with respect to etch time for the samples annealed at (a) 820°C, (b) 800°C, (c) 780°C, (d) 750°C, (e) 700°C and (f) 650°C, respectively. The core level binding energy spectra for these six representative samples are presented in Figure 3 &4 and in Figure S1 in the Supplementary Material. The black dashed lines at 150 sec and 360 sec indicate the position related to the middle of the film and the interface, respectively. The two mentioned positions are relevant for the next sections where the spectra deconvolution and quantitative analysis are shown.

**Fig. 6:** Comparison of elemental concentration individually among the representative six samples. The atomic percentage of (a) oxygen, (b) nitrogen and (c) silicon among the samples are compared with etching time. The comparison of core level binding energy spectra, obtained after 150 sec etching, for (d) Ti 2p, (e) N 1s and (f) Si 2p among the samples. In order to compare the related XPS-spectra from the film only, etching time of 150 sec is selected as that provides the information mostly from the film and is shown by the black dashed vertical line in Figure 5. The intensities for the individual curves in (d)-(f) are normalized with respect to the highest intensities in their respective scans and the curves are shifted upward for clarity.

**Fig. 7:** The core level binding energy spectra for Ti 2p (left column), N 1s (middle column) and Si 2p (right column), obtained after 150 sec of etching for all the six representative samples. Individual samples representing various annealing temperatures are placed in row-

wise. Different phases are determined from the fitted curves as marked in the figure. The details are explained in the text.

**Fig. 8:** Comparison of nitride and silicide phases as obtained from the fitting. (a)-(f) The relative concentration (%) of nitride and silicide phases with etching time for the representative six samples annealed at 820°C, 800°C, 780°C, 750°C, 700°C and 650°C, respectively. Scattering points are obtained from the fitting as explained in the text and the line segments are the guide to eye.

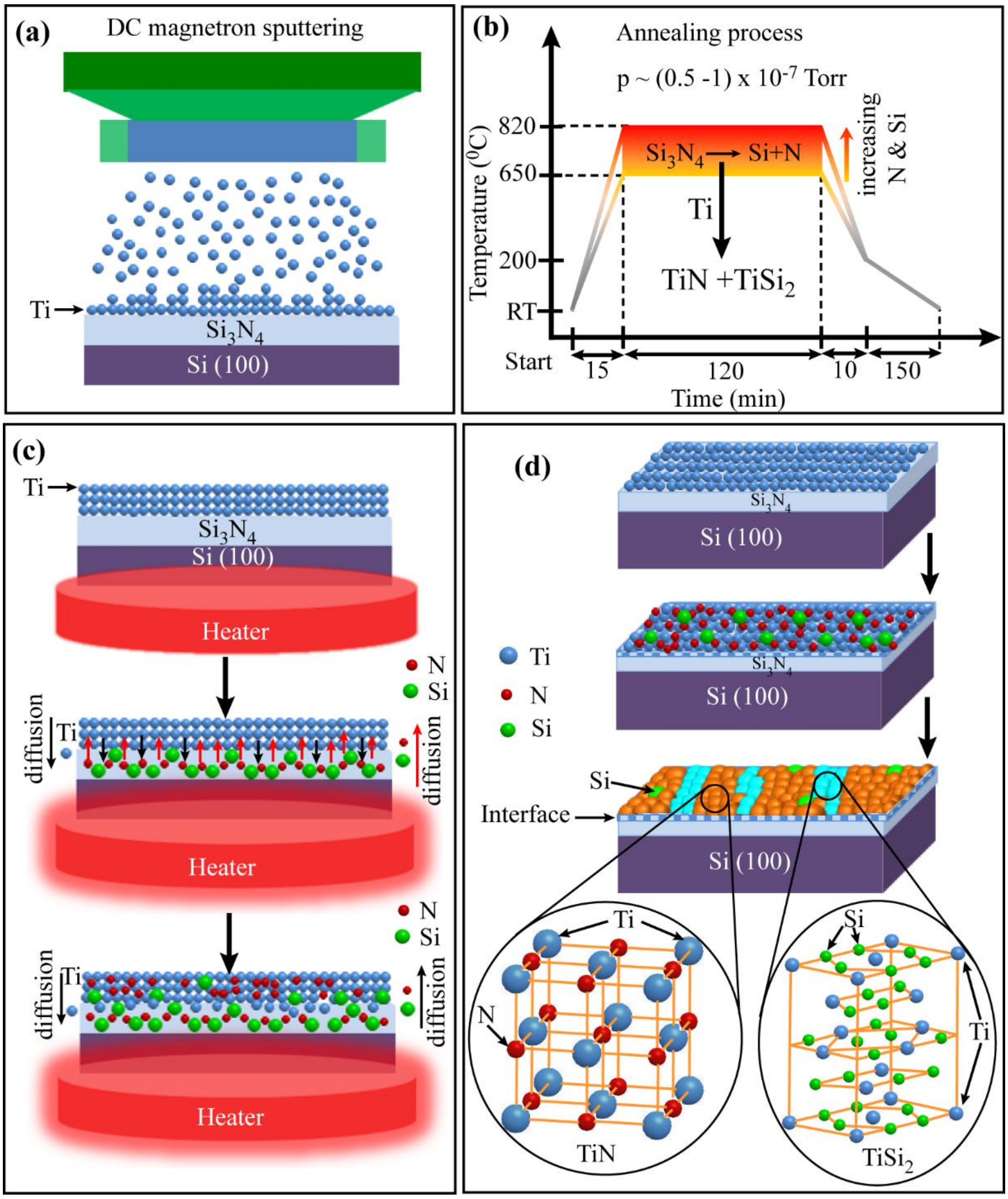

**Fig. 1**

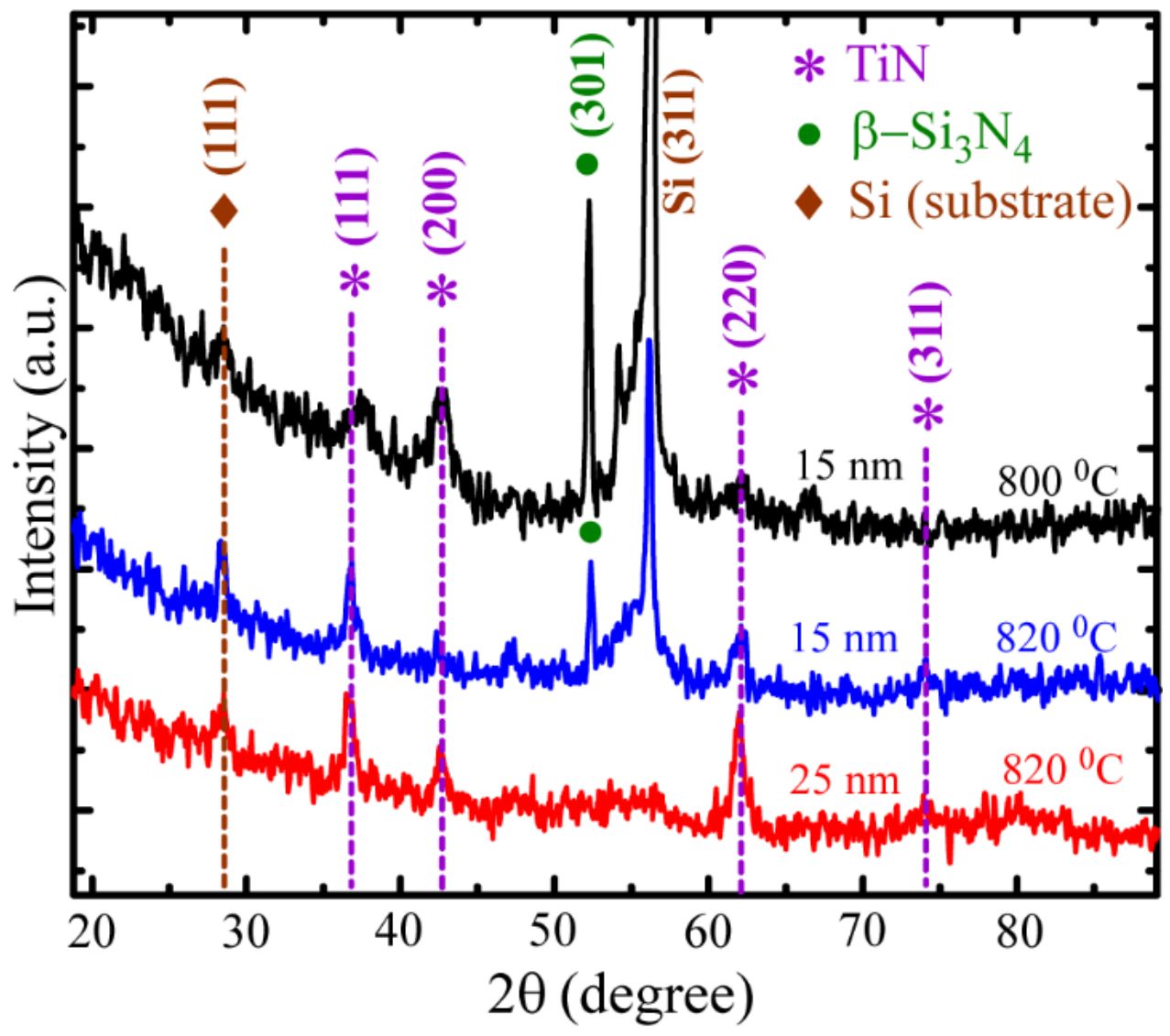

**Fig. 2**

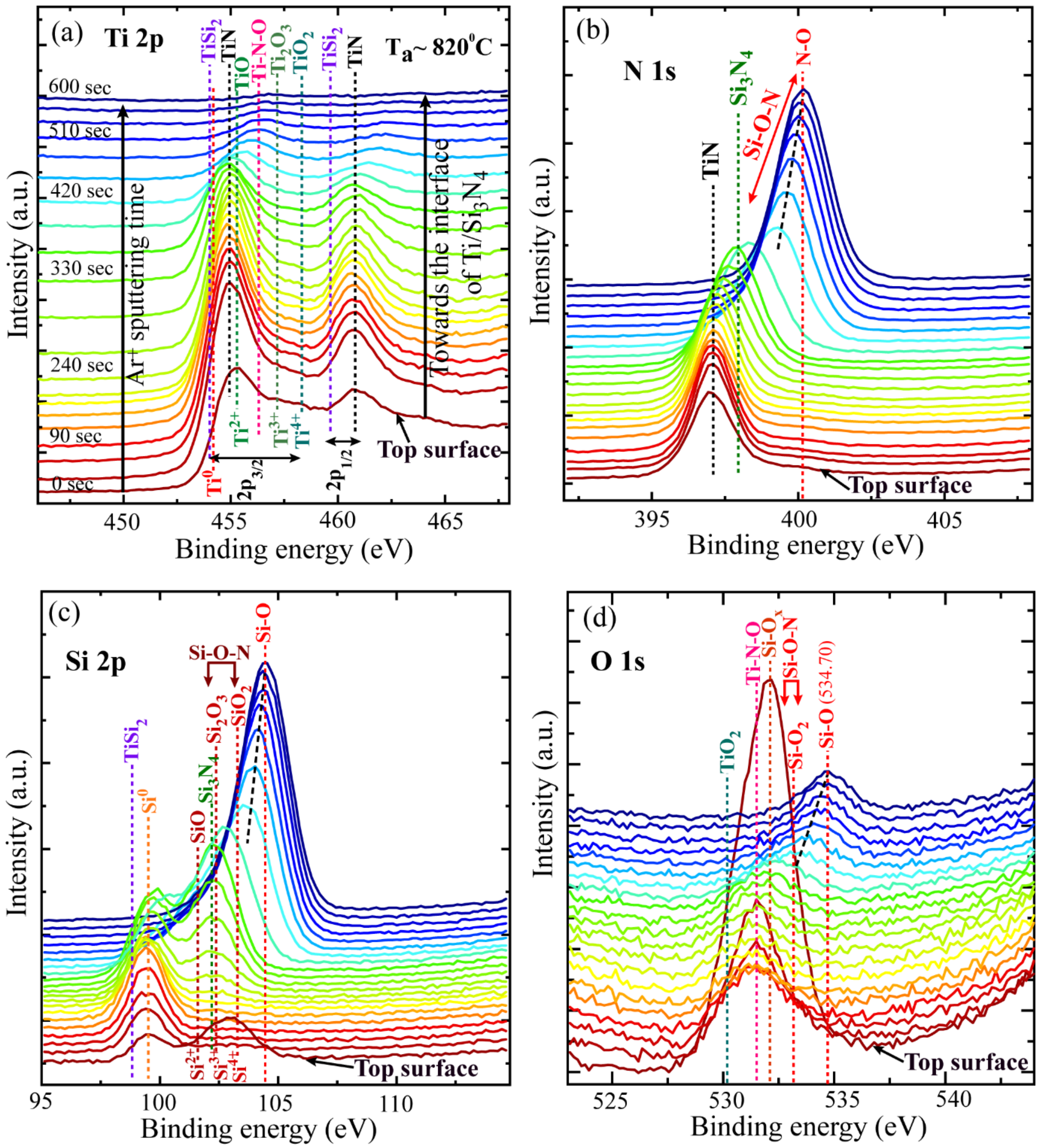

**Fig. 3**

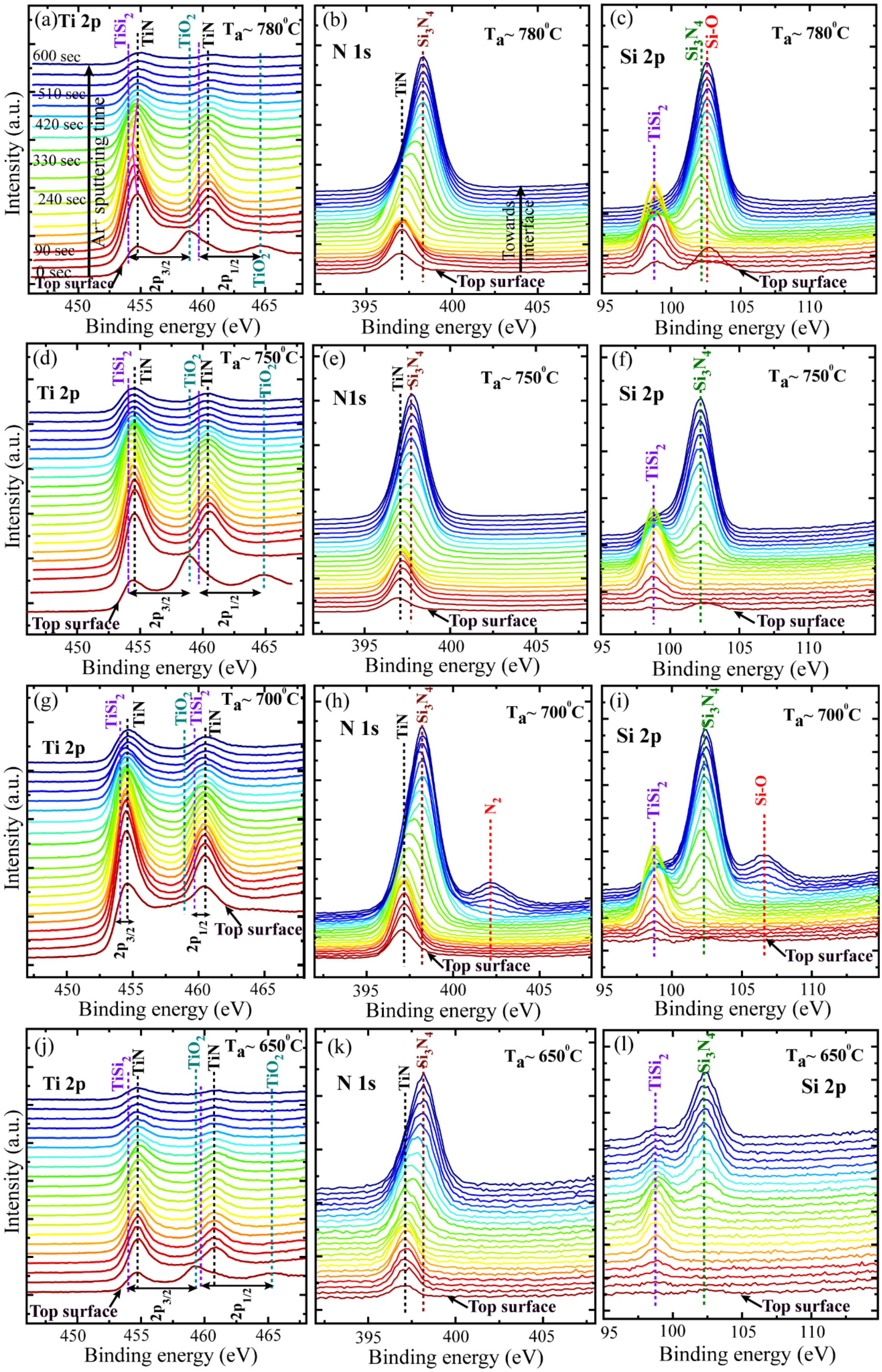

**Fig. 4**

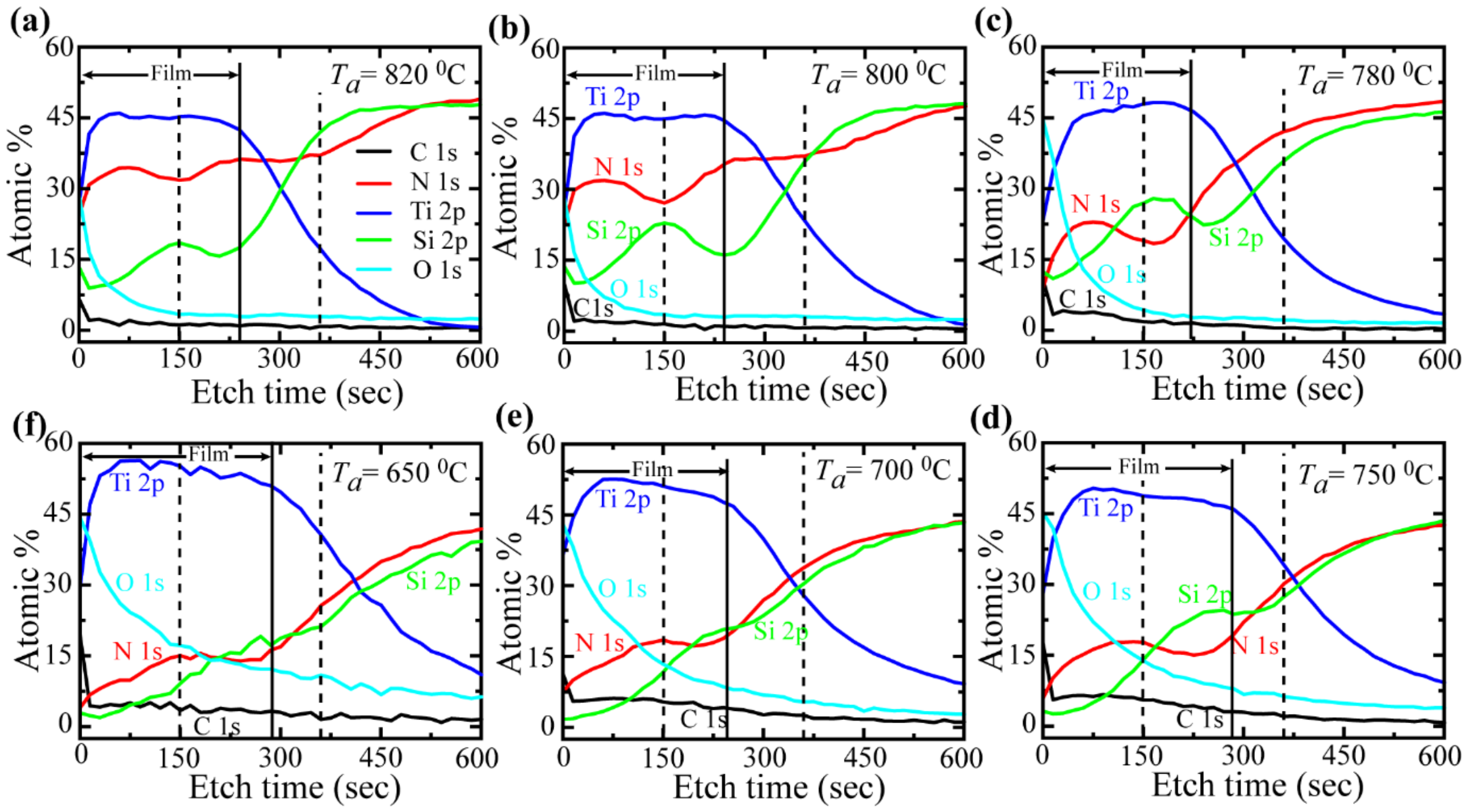

**Fig. 5**

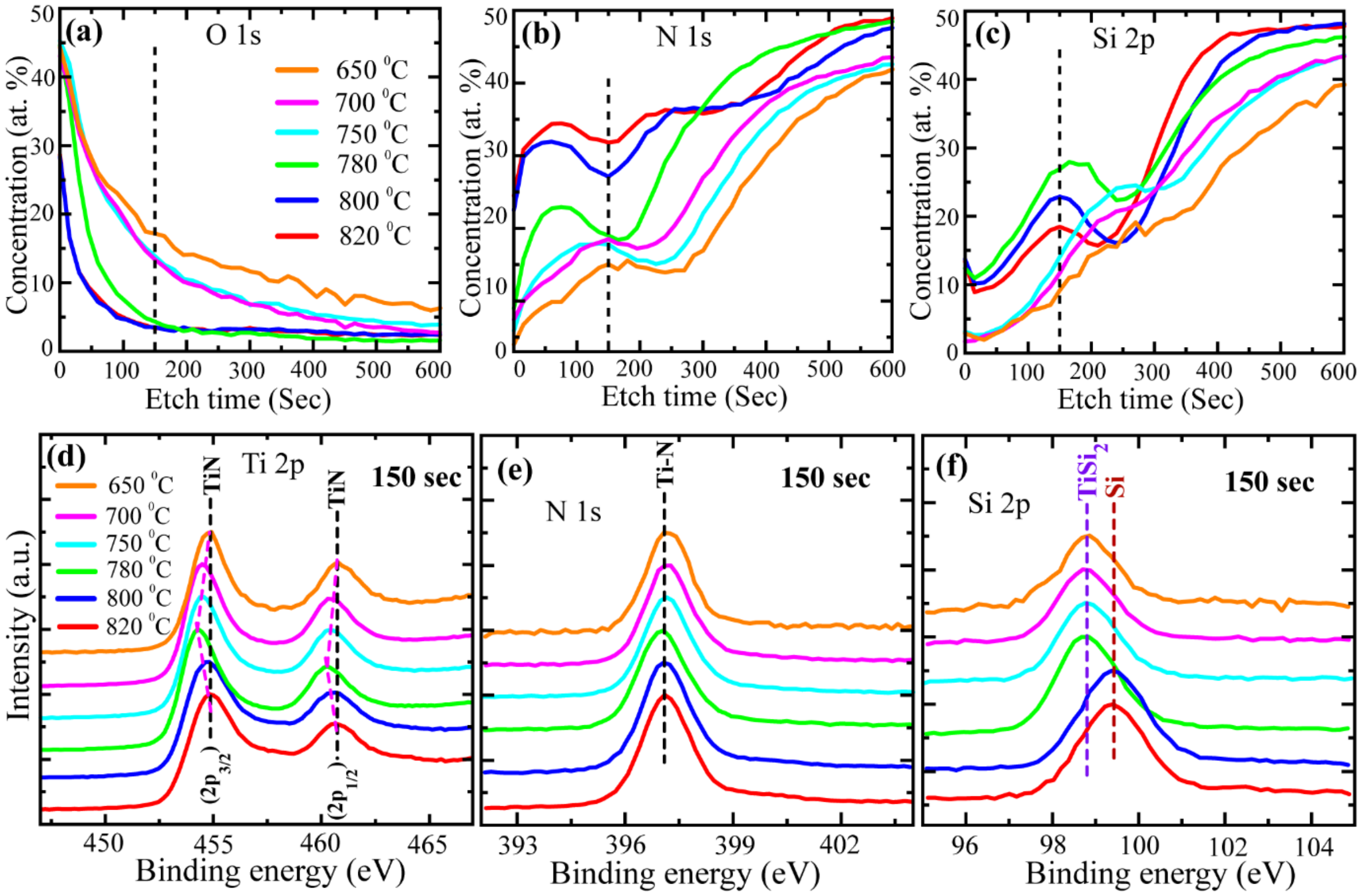

**Fig. 6**

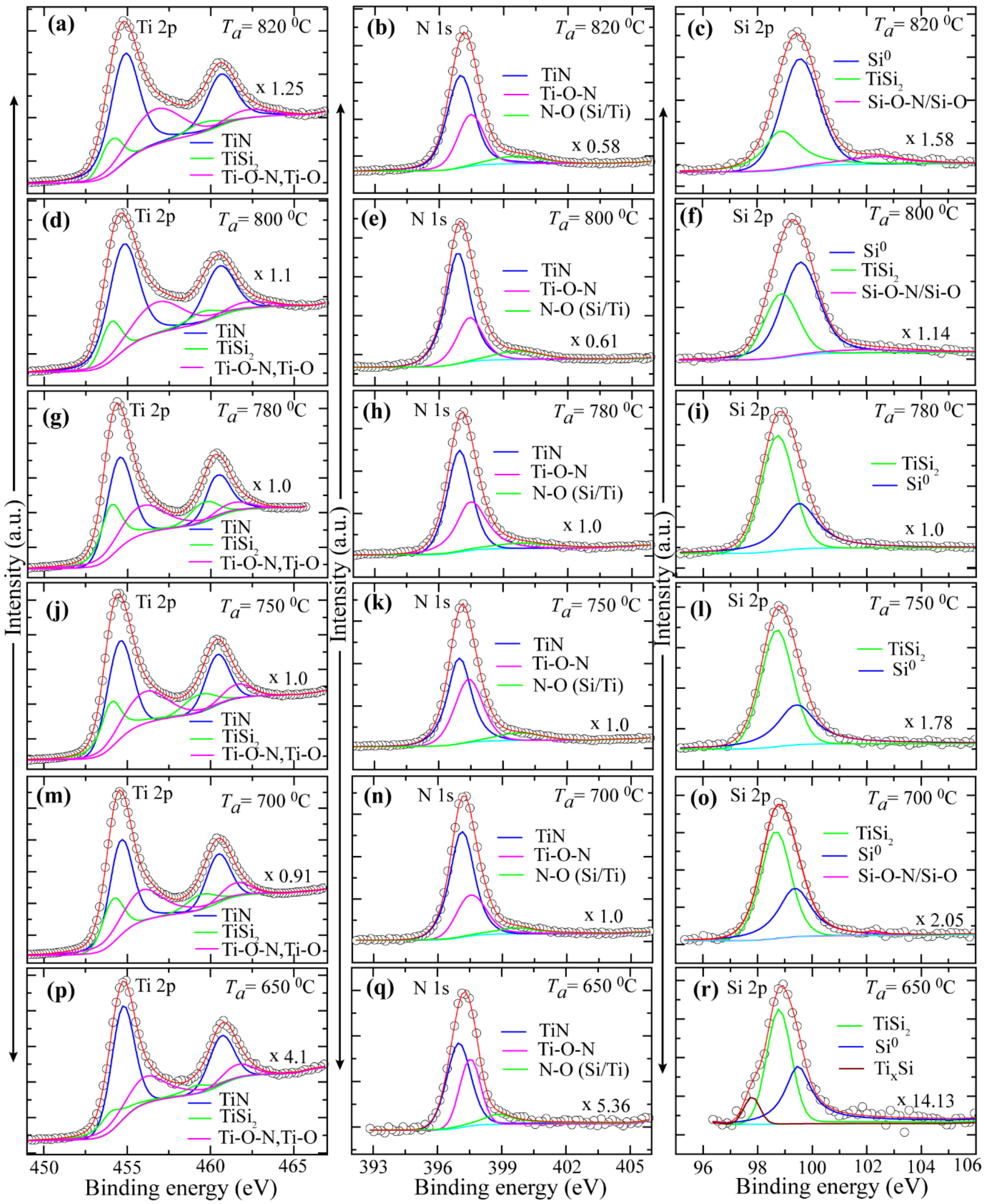
(a) Ti 2p $T_a$= 820 $^0$C
x 1.25
TiN
$TiSi_2$
Ti-O-N,Ti-O
(b) N 1s $T_a$= 820 $^0$C
TiN
Ti-O-N
N-O (Si/Ti)
x 0.58
(c) Si 2p $T_a$= 820 $^0$C
$Si^0$
$TiSi_2$
Si-O-N/Si-O
x 1.58
(d) Ti 2p $T_a$= 800 $^0$C
x 1.1
(e) N 1s $T_a$= 800 $^0$C
x 0.61
(f) Si 2p $T_a$= 800 $^0$C
x 1.14
(g) Ti 2p $T_a$= 780 $^0$C
x 1.0
(h) N 1s $T_a$= 780 $^0$C
x 1.0
(i) Si 2p $T_a$= 780 $^0$C
x 1.0
(j) Ti 2p $T_a$= 750 $^0$C
x 1.0
(k) N 1s $T_a$= 750 $^0$C
x 1.0
(l) Si 2p $T_a$= 750 $^0$C
x 1.78
(m) Ti 2p $T_a$= 700 $^0$C
x 0.91
(n) N 1s $T_a$= 700 $^0$C
x 1.0
(o) Si 2p $T_a$= 700 $^0$C
x 2.05
(p) Ti 2p $T_a$= 650 $^0$C
x 4.1
(q) N 1s $T_a$= 650 $^0$C
x 5.36
(r) Si 2p $T_a$= 650 $^0$C
$Ti_xSi$
x 14.13
Intensity (a.u.)
Binding energy (eV)
450
455
460
465
393
396
399
402
405
96
98
100
102
104
106

**Fig. 7**

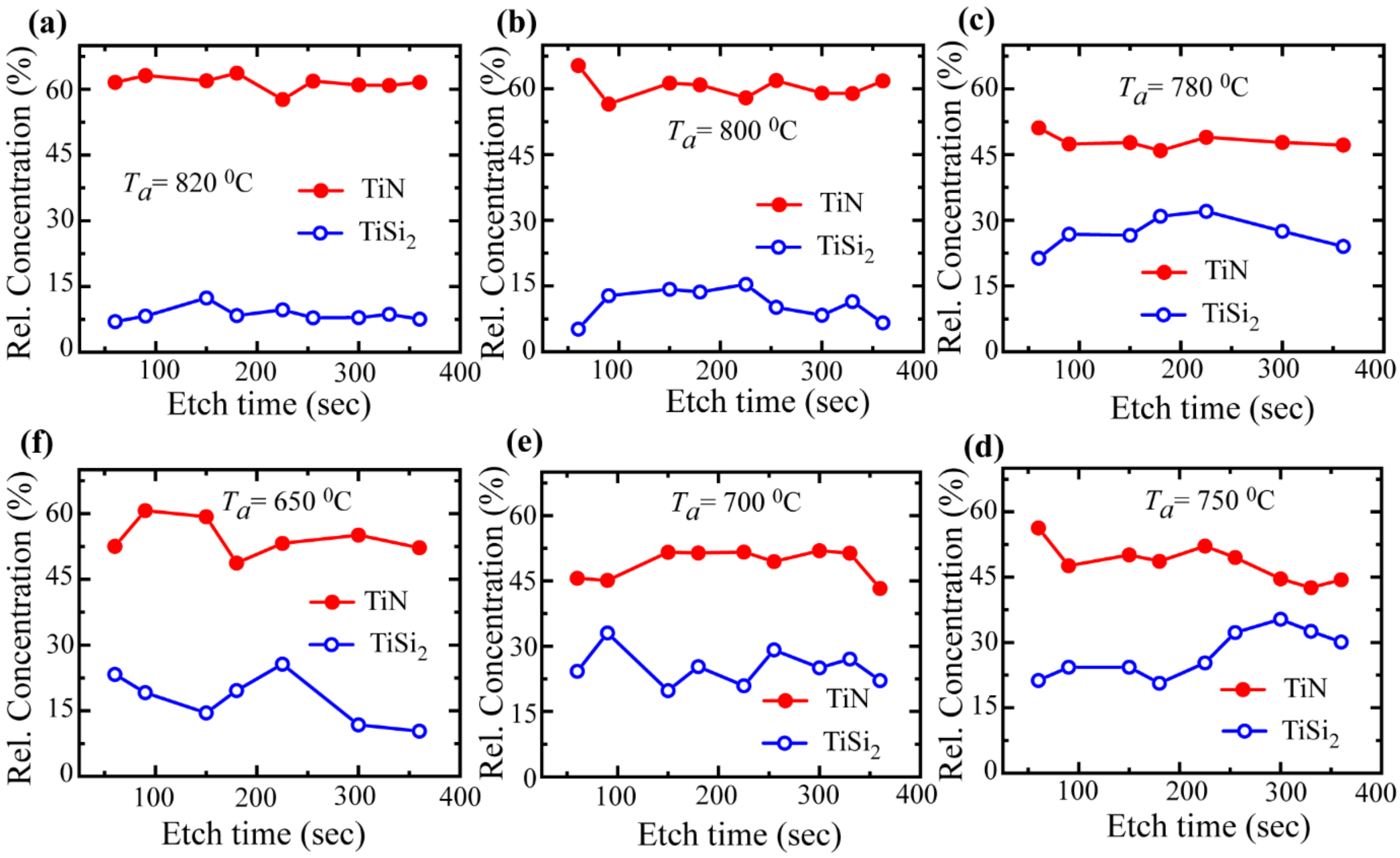

**Fig. 8**

Table 1: Comparison of the binding energy obtained from the fitting used in Fig. 7 with literature values

| Chemical state | Binding Energy (eV) | | | | Reference | Note |
|---|---|---|---|---|---|---|
| | Ti $2p_{3/2}$ | N 1s | Si 2p | O 1s | | |
| TiN | 454.55 | 396.8 | | | [70] | |
| | 454.95 - 455.2 | 397.11 - 397.21 | | | [40], [45], [63] | |
| | 454.8 | 396.8 | | | [71] | |
| | 454.7 | 396.3 | | | [44] | |
| | 455.1 | 396.9 | | | [46] | |
| | 454.7 | | | | [62] | |
| | 454.5-454.8 | 396.9-397.1 | | | This work | (a) |
| $TiSi_2$ | 453.9 | | | | [43] | |
| | 454.1 | | 99.3 | | [49] | |
| | 453.7 | | 98.6 | | [59] | |
| | 454 - 454.1 | | 98.6-98.9 | | This work | |
| Ti-N-O | 457 | 397.7 | | 531.2 | [14] | |
| | 455.1 | 397.5-397.7 | | 531.6-531.9 | [61], [68] | |
| | 455.9 | 399.3 | | 531.4 | [71] | |
| | 456.7 | 398 | | 531.8 | [44] | |
| | 456.9 | | | 531.8 | [48] | |
| | 456.3 | 395.8 | | | [62] | |
| | 455.5 | 396.1 | | 531.7 | [46] | |
| | 456.6 | 396.17 | | 530.99 | [65] | |
| | 455.8 - 456.8 | 397.4-397.5 | | 531.5 | This work | |
| N-O | | 399.46 | | 532.99 | [40] | |
| | 457.4 | 400.5 | | | [63] | |
| | | 400 | | | [69] | |
| Si | | | 99.6 | | [49] | |
| | | | 99.5 | | [60] | |
| | | | 99.5 | | [50] | |
| | | | 99.3-99.6 | | This work | |
| Si-O-N | | 400.7 | | | [70] | |
| | | | 102.5 | | [51] | |
| | | 398.5 & 400 | 103.4 | | [54] | |
| | | | 102.6 | 532.8 | [53] | |
| | | 399.6-399.8 & 398.8 | 101.9-102.3 | | This work | (c) |
| $TiO_2$ | 458.3 | | | 530.3 | [71] | |
| | 457.9 | | | 530.0 | [44] | |
| | 458.7 | | | 529.6 | [48] | |
| | 459.2 | | | 530.2 | [63] | |
| | 458.8 | | | | [62] | |
| | 458.2 | | | 530 | [58] | |
| | 458.8 | | | 530.2 | This work | |
| $SiO_2$ | | | 103.8 | | [54] | |
| | | | 103.3 | 533.2 | [53] | |
| | | | 103.5 | 533 | [58] | |
| $SiO_x$ | | | 102.8 | 532.1 | This work | (b) |
| $Si_3N_4$ | | 398.3 | 100.8 | | [48] | |
| | | 397.7 | 102.5 | | [54] | |
| | | | 102 | | [51] | |
| | | 398.1 | 101.6 | | [53] | |
| | | 398.2 | 102.3 | | This work | |

(a) The range covers all the studied samples with annealing temperature ranging from 650℃-820℃
(b) On the top surface for $T_a$ = 800℃, 820℃
(c) 398.8 corresponds to sample annealed at 650℃

## Supplementary Material

**Interface study of thermally driven chemical kinetics involved in Ti/$Si_3N_4$ based metal-substrate assembly by X-ray photoelectron spectroscopy**

Sachin Yadav[a,b] and Sangeeta Sahoo[a,b,*]

[a]*Academy of Scientific and Innovative Research (AcSIR), AcSIR Headquarters CSIR-HRDC Campus, Ghaziabad, Uttar Pradesh 201002, India.*

[b]*Electrical & Electronics Metrology Division, National Physical Laboratory, Council of Scientific and Industrial Research, Dr. K. S. Krishnan Marg, New Delhi-110012, India.*

[*]Corresponding author.

*E-mail address:* sahoos@nplindia.org (Sangeeta Sahoo)

❖ **X-ray photoelectron spectroscopy (XPS) spectra for a sample annealed at 800°C**

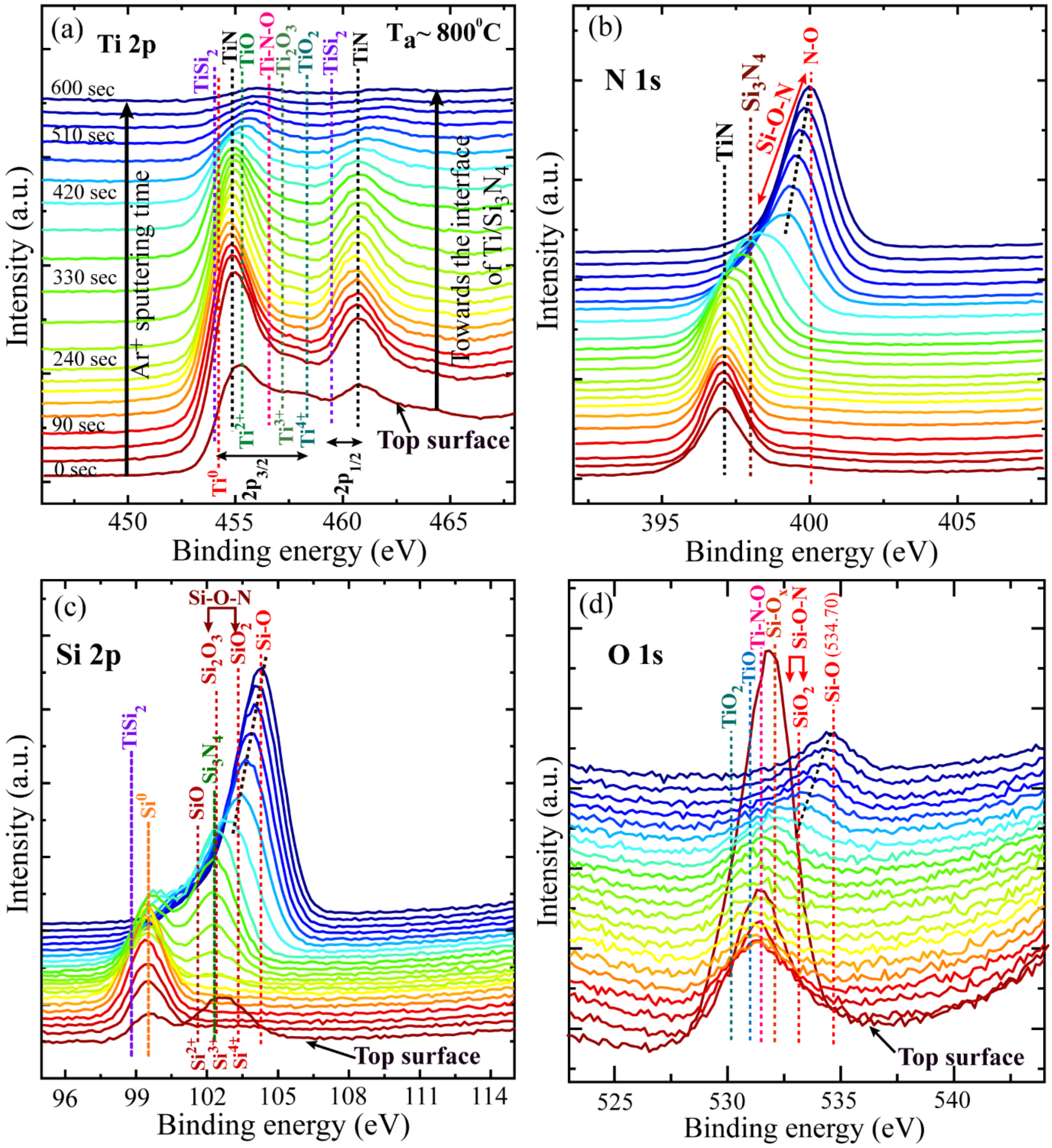

**Fig. S1:** The core level binding energy XPS spectra of Ti 2p (a), N 1s (b), Si 2p (c) and O 1s (d) for a representative sample of 15 nm thickness and annealed at 800°C for 2 hours. The XPS spectrum measured on the top surface of as-loaded sample is shown at the bottom and in the upward direction, the curves represent the scans measured after etching the film (in steps with constant time and energy) from the top surface till the substrate $Si_3N_4$/Si is reached.

In Fig, S1, we present the XPS spectra collected from a sample annealed at 800°C. Similar to the samples presented in the main manuscript in Figs. 3&4, the sample was etched in situ by $Ar^+$ ions with 500 eV energy for a fixed period of time (15 sec) in steps and after each etching step the spectra were measured. Here, we have presented the spectra in 30 sec intervals of etching. The bottom curve represents the top surface for the as-loaded sample before etching started. Towards the up direction, the spectra represent the XPS data after consecutive etching until the substrate is reached. The binding energy positions for TiN (454.9 eV), $TiSi_2$ (454 eV) and Ti-O-N (456.5 eV) are shown in Fig. S1(a) for Ti 2p. Further, different oxidation states of Ti are also marked in Ti2p spectra with binding energies such as $Ti^{2+}$ (454.4 eV), $Ti^{3+}$ (457.2 eV), $Ti^{4+}$ (458.8 eV). Oxide formation on the top surface is evident by the presence of a broad peak at around the binding energy corresponding to $TiO_2$. However, TiN shares the strongest peak while $TiSi_2$ appears at the shoulder of the peak. The presence of TiN is strongly supported by N 1s spectra presented in Fig. S1(b). Whereas, Si 2p spectra reveals the dominance of elemental Si over the silicide phase. At the interface the presence of $Si_3N_4$ is evident in both N 1s spectra and Si 2p spectra. However, further down through the $Si_3N_4$ layer, the shifting of the peaks towards higher binding energy indicates the formation of silicon oxynitride (Si-O-N) in the region lying between $Si_3N_4$ & N-O in N1s and between $Si_3N_4$ & $SiO_2$ in Si2p as shown in Fig. S1(b) & (c), respectively. We have observed further shift in binding energy towards higher energy due to the formation of silicon oxide (Si-O), close to the Si substrate, as shown with red dotted line in Fig. S1(c). Moreover, the formation of Ti-N-O (531.5 eV), oxides of Ti [$TiO_2$ (530.0 eV) &TiO (531.0 eV)] and oxides of silicon [Si-$O_x$ (532.10 eV), SiO2 (532.8 eV) & Si-O (534.7 eV)] are also confirmed by O 1s as shown in Fig S1(d). Finally, the XPS analysis of the sample annealed at 800°C (Fig.

S1) illustrates the similar results obtained for the sample annealed at 820°C as presented in Fig. 3 in the main manuscript.

- **O 1s spectra for the samples annealed at 780°C, 750°C, 700°C and 650°C.**

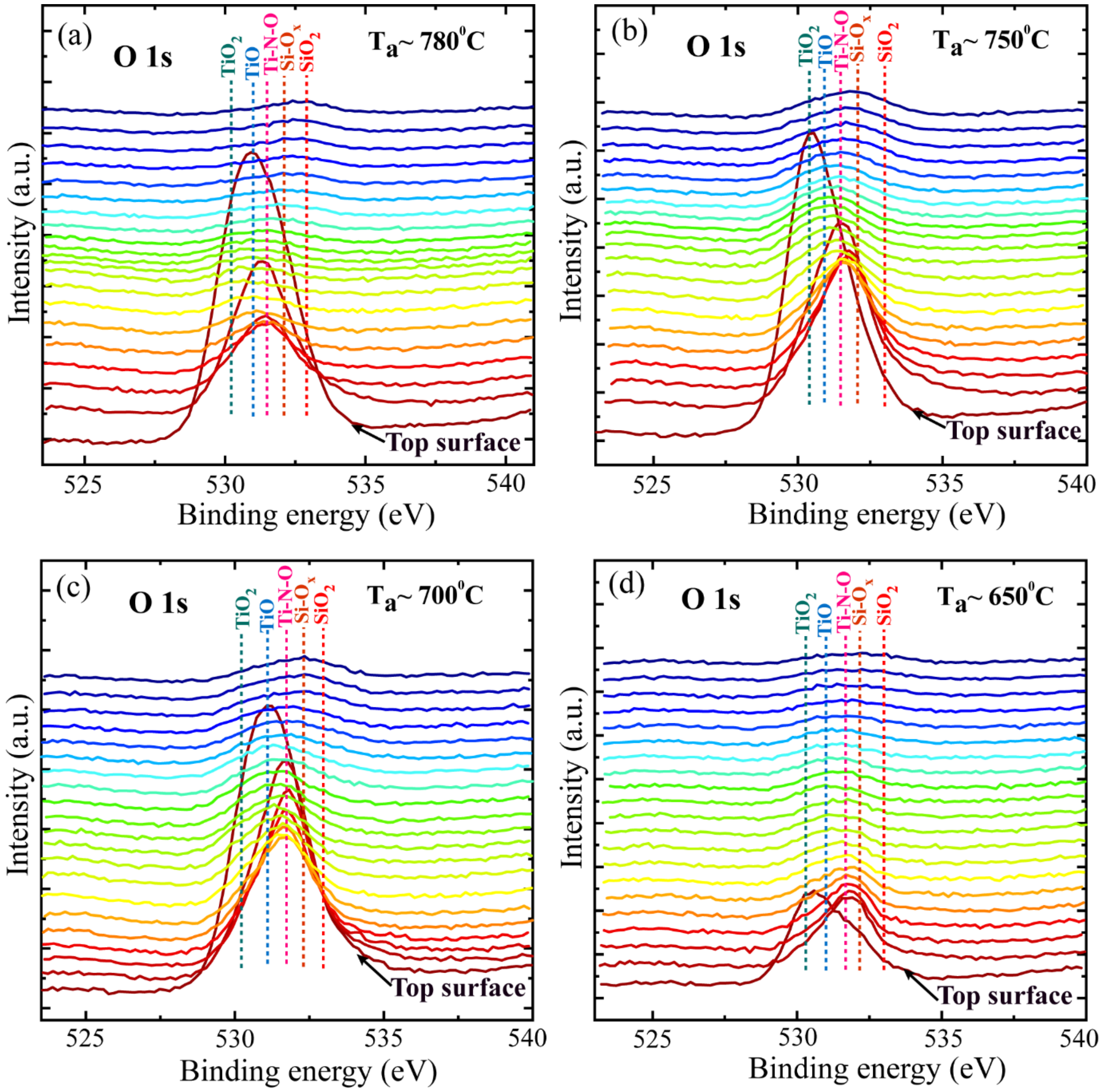

**Fig. S2:** The core level binding energy spectra of O1s for the samples shown in Fig. 4 in the main manuscript. The annealing temperature, Ta, corresponds to (a) 780 °C, (b) 750 °C, (c) 700 °C, and (d) 650 °C. The thickness of Ti film is 15 nm for all the samples presented here. Samples were etched out by using $Ar^+$ ions sputtering for depth profile study.

In Fig. S2, we have represented O 1s spectra for the samples which were annealed at 780°C and below. Ti 2p, N 1s and Si 2p spectra for this set of samples are shown in Fig. 4 in the main manuscript. The Top surface scan for all samples are having two major peaks that are TiO (531.0 eV) & $TiO_2$ (530.0 eV) accompanied by minor contribution from the other oxide peaks those are Ti-N-O (531.5), Si-Ox (532.1) & SiO2 (532.8) as shown in Fig. S2. On further moving towards the interface, we have observed increment in the intensity of minor oxides peaks and decrement in the intensity of major peaks due to decrease in the oxygen concentration shown in Fig. 5 in the main manuscript.

- **Survey scan for all the annealed samples presented in this study and one control sample without annealing**

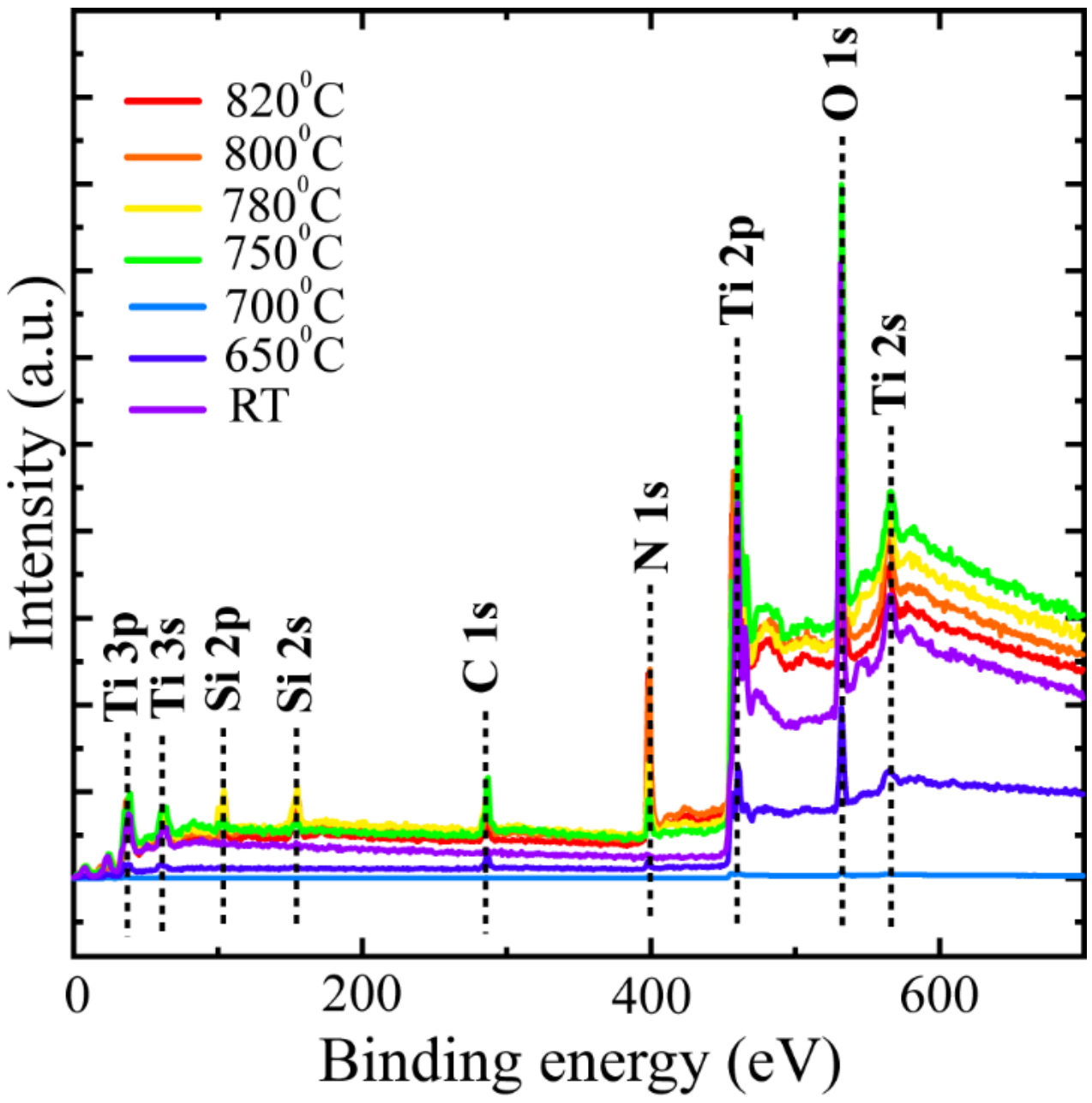

**Fig. S3:** XPS survey spectra for all the annealed samples except for the one which represents the room temperature (RT) grown Ti film without any annealing.

In this figure, we have represented the survey spectra of all the annealed samples along with one sample that was not annealed after the Ti deposition. The main peaks in the survey spectra are O1s, Ti2p, N1s, C1s, and Si2p. Survey spectra show the change in the intensity with respect to the annealing temperature. Here, it should be pointed out that the intensities for the samples (700°C) & (650°C) are much less with respect to the intensity of other spectra.

❖ **X-ray photoelectron spectroscopy (XPS) spectra for a sample without annealing**

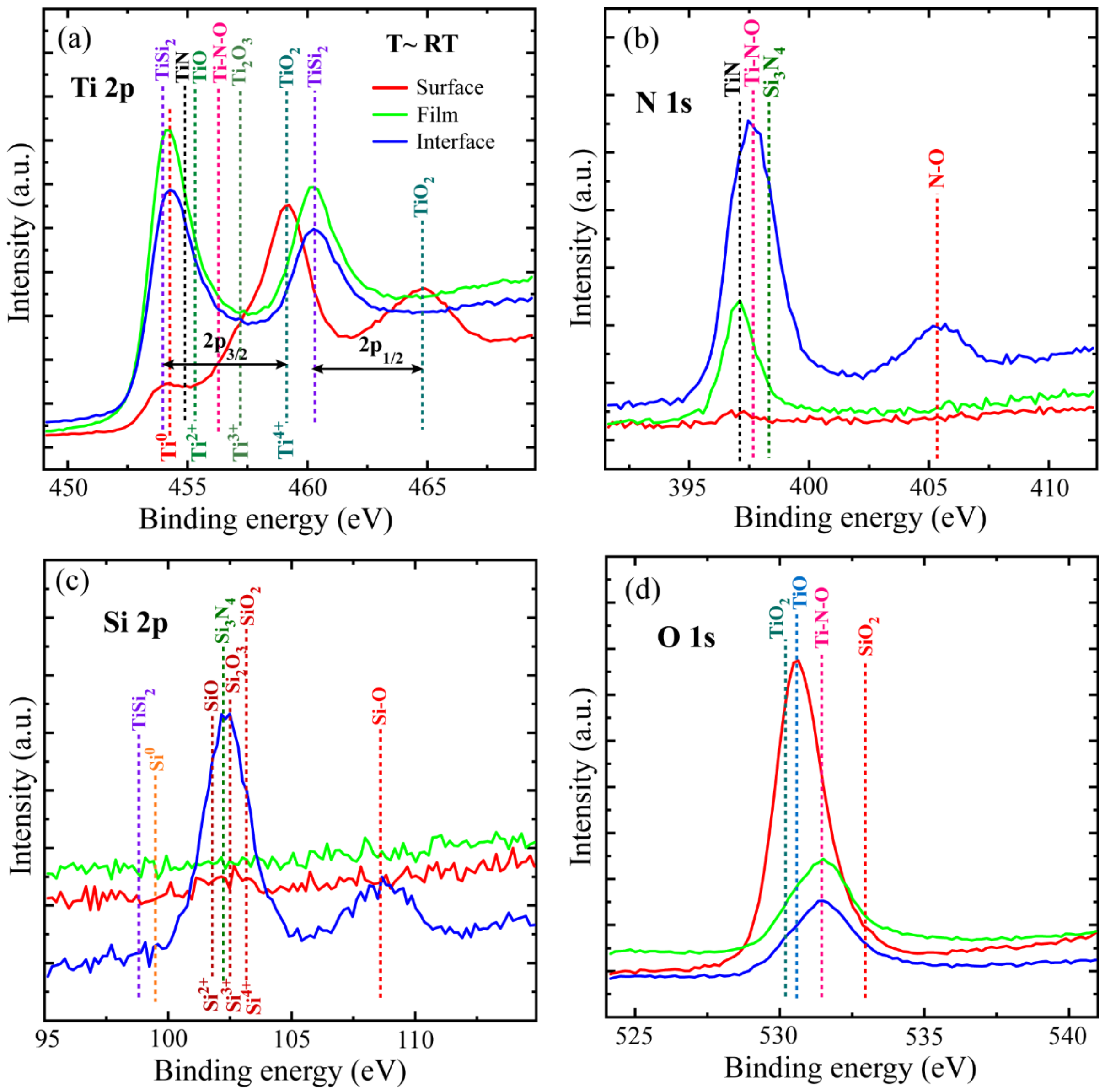

**Fig. S4:** The core level binding enery spectra for the sample (15 nm Ti film) that was grown at room temperature without any annealing. It consists of spectra for Ti2p (a), N1s (b), Si 2p (c) and O1s (d). Here, the scans are selected from three different regions, namely, on the surface, inside the film and at the interface.

In order to compare the results with annealed samples, the XPS core binding energy spectra presented in Fig. S4 are taken from the sample that was not annealed after Ti deposition. For comparison, we have taken three scans from different regions that include the top surface (red), inside the film (green) and at the interface (blue). The surface scan in Ti 2p spectrum shows the dominance of $TiO_2$ peak at (458.8 eV) along with a board peak positioned at 454.2 eV marking the presence of elemental Ti. Further, as we go inside the film, we found the dominance of elemental Ti peak over the $TiO_2$ peak along with the emergence of low-intensity TiN peak at 454.9 eV. Here, it should be noted that even for the sample, which was not annealed, TiN peak appears. This indicates that due to possible heat generation during Ar+ sputtering, a small amount of N can come out from $Si_3N_4$ layer and react with Ti. However, there is no Si detected in the middle of the sample. For the annealed samples, we do see the dominance of TiN peak at 454.9 eV along with minor contribution from the elemental Ti peak as shown in Fig. 3(a) & 4 in the main manuscript and Fig. S1(a) in the supplementary. As we reach at the interface, dominance of elemental titanium peak starts to decrease with increase in the concentration of TiN peak due to the presence of more N atoms near the interface. However, presence of TiN is also confirmed by N1s spectrum, where TiN peak at 397.0 eV starts to increase inside the film along with oxynitride peak positioned at 397.7 eV till the interface as shown in Fig. S4(b). However, top surface scan reveals that there is very little intensity related to the TiN peak marked in Fig. S4(b). Interestingly, during the course of study, we found that silicon doesn’t react with titanium to form either $TiSi_2$ or present in the elemental form on the surface or inside the film as shown in the Fig. S4 (c). Whereas, in the case of annealed samples, we found that silicon starts to form $TiSi_2$ or remains in the elemental form on the top surface as well inside the film as shown in the Fig. 3(c) & 4 in the main manuscript and Fig. S1(c) in the supplementary. Further moving towards the interface, Si 2p spectrum starts to show $Si_3N_4$ peak at 102.0 eV and Si-O peak at (103.2 eV) in the Fig. S1(c), presence of Si-O may be due to the decomposition of Si3N4 at the interface by $Ar^+$ ions sputtering. Moreover, O1s spectrum represents that top surface is being dominated by TiO (531.0 eV) and $TiO_2$ (530.0 eV) as sample was being exposed to the air before loading into the system for XPS analysis. Further, as we move inside the film, oxide peaks of Ti is being reduced, while oxynitride (Ti-N-O) peak positioned at 531.5 eV starts to dominate due to the presence of TiN as minority phase.

❖ **N 1s spectra and the atomic profile for the sample annealed at 780 °C for an extended period of etching**

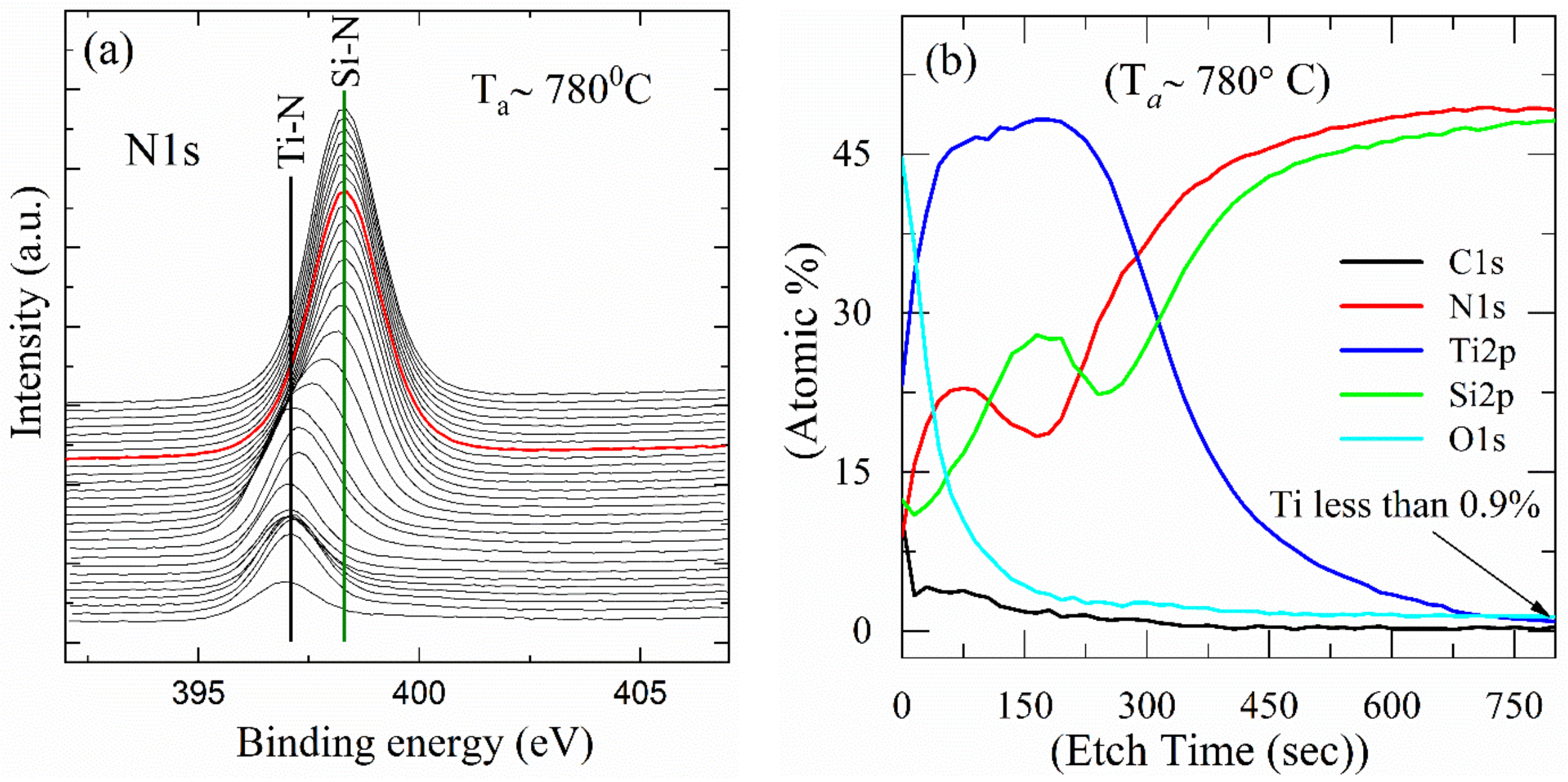

Fig. S5: (a) XPS core level binding energy spectra of N1s for the sample annealed at 780°C. The spectrum represented in red colour represents the scan taken after 600 sec sputtering time. (b) Atomic concentration depth profile for the same sample is shown.

The XPS core level binding energy spectra of N 1s for the sample that was annealed at 780°C is presented in Fig. S5(a), where first scan from the bottom represents the top surface scan and thereafter followed up by the rest of the scans as we move towards the substrate. The red coloured scan was taken after 600 sec of sputtering time and we have already shown N 1s spectra up to this time interval in Fig. 4(b) in the main manuscript. However, in this spectra, we have presented up to 810 sec of sputtering time and at this point of time, the atomic concentration of Ti becomes less than 0.9%, as shown in Fig. S5(b) [marked with a black solid arrow]. Previously, we have observed shift in binding energy towards the higher side for the samples annealed at 800°C and above near the interface [as shown in Fig. 3 and Fig. S1], but we didn't observe any such shifting for the samples annealed at 780°C as shown in Fig. S5(a). In order to rule out the effect of charging on peak shifting, we have taken a close look at Ti concentration present after 600 sec in the depth profile for 820°C (800°C) & 780°C and found that for 820°C ( 800°C), the concentration of Ti after 600 sec is ~ 0.9% (~ 1.5%) and for 780°C, it is close to 3.5%. At the time just after 600 sec of sputtering, as the percentage of Ti is very less for the sample annealed at 820°C compared to that of the sample annealed at 780°C, the charging can cause the binding energy shifting for the case of 800°C. And it is justified that there is very little charging effect for the sample annealed at 780°C and hence, no shifting is observed. Therefore, in order to see the effect of charging on binding energy shifting for 780°C, we have etched the sample further [*in-situ*] till the concentration of Ti reaches to less than 1%. However, further etching for the sample annealed at 780°C, Ti concentration gets reduced and after 800 sec of sputtering, the same reaches to ~ 09%%, as marked with black solid arrow in Fig. S5(b). Here, the charging effect should lead to the binding energy shifting, but, we do not observe any such shifting in N1s spectra as shown in Fig. S5(a). If the presence of Ti after 600 sec in 780°C compensates the charging effect, then the shifting would appear when the concentration of Ti is less. Thus, we conclude that the charging effect is not playing a significant role in shifting the peaks to higher binding energy at the interface region.

- **Tabular representation of the fitting parameters used for the fitting shown in Fig. 7 of the main manuscript**

Table S1 presents the fitting parameters used to fit the XPS spectra obtained after 150 sec etching with $Ar^+$ ion for all the representative six samples annealed at six different temperatures between 650°C to 820°C. As explained in the manuscript, the fittings were done by using mixed Lorentzian/Gaussian based peak fitting functions in the framework of Avantage Software from ThermoFisher. Shirley background correction was used for the background subtraction for all the elements. While fitting, we have taken into consideration the constraints related to spin splitting binding energy gap related to Ti $2p_{3/2}$ and $2p_{1/2}$ and their area ratio for TiN and $TiSi_2$. While, the oxygenated part (Ti-N-O) is combined with a single wide peak containing the oxynitride as well as any possible oxide phase. The area ratio for Ti $2p_{3/2}$ and $2p_{1/2}$ was maintained at close to 2:1 and the splitting energy gap for TiN was varying in the range between 5.7 - 6.0 eV for the fittings used to obtain Fig. 8 in the article. For TiSi2, the splitting binding energy varied between 5.4 eV to 5.7 eV. Here, the fitting parameters include the peak position of the binding energy, full-width-at-half-maximum (FWHM), and the peak area for Ti 2p, N 1s and Si 2p. For Ti 2p, the peak area represents the sum of the peak areas for both the line shapes Ti $2p_{3/2}$ and $2p_{1/2}$.

**Table S1: Fitting parameters used for the fittings in Fig. 7 in the main manuscript**

| Sample ($T_a$ in °C ) | Ti $2p_{3/2}$ | | | N 1s | | | Si 2p | | |
|---|---|---|---|---|---|---|---|---|---|
| | **BE (eV)** | **FWHM** | ***at. %** | BE (eV) | **FWHM** | **at. %** | **BE (eV)** | **FWHM** | **at. %** |
| **(820 °C )** | 454.84 TiN | 2.04 | 61.95 | 396.99 TiN | 1.49 | 60.29 | 99.52 $Si^0$ | 1.62 | 64.99 |
| | 454.0 $TiSi_2$ | 1.54 | 12.37 | 397.47; Ti-O-N | 1.45 | 31.32 | 98.8 $TiSi_2$ | 1.57 | 26.68 |
| | 456.7 Ti-O-N, Ti-O | 3.5 | 18.4 | 399.6 N-O (Si/Ti) | 2.77 | 8.39 | 102.32 Si-O-N/Si-O | 2.34 | 8.33 |
| **(800 °C )** | 454.75 TiN | 2.13 | 61.34 | 396.86; TiN | 1.35 | 64.14 | 99.59 $Si^0$ | 1.6 | 60.63 |
| | 454.0 $TiSi_2$ | 1.42 | 14.27 | 397.41; Ti-O-N | 1.42 | 26.68 | 98.89 $TiSi_2$ | 1.46 | 33.51 |
| | 456.79 Ti-O-N, Ti-O | 3.5 | 24.4 | 399.66 N-O (Si/Ti) | 2.59 | 9.18 | 101.89 Si-O-N/Si-O | 4.07 | 5.85 |
| **(780 °C )** | 454.48 TiN | 1.78 | 47.73 | 396.95 TiN | 1.35 | 55.03 | 98.73 $TiSi_2$ | 1.39 | 62.87 |
| | 454.0 $TiSi_2$ | 1.51 | 26.57 | 397.43 Ti-O-N | 1.66 | 36.06 | 99.51 $Si^0$ | 1.6 | 37.13 |
| | 455.8 Ti-O-N, Ti-O | 3.2 | 25.7 | 399.7 N-O (Si/Ti) | 3.57 | 8.92 | | | |
| **(750 °C )** | 454.55 TiN | 1.78 | 50.12 | 396.93 TiN | 1.31 | 51.14 | 98.69 $TiSi_2$ | 1.38 | 63.04 |
| | 454.0 $TiSi_2$ | 1.48 | 24.32 | 397.40 Ti-O-N | 1.53 | 41.29 | 99.38 $Si^0$ | 1.68 | 36.96 |
| | 456.09 Ti-O-N, Ti-O | 3.07 | 25.58 | 399.76 N-O (Si/Ti) | 2.5 | 7.57 | | | |
| **(700 °C )** | 454.59 TiN | 1.65 | 51.58 | 397.08 TiN | 1.35 | 62.43 | 98.65 $TiSi_2$ | 1.29 | 59.01 |
| | 454.12 $TiSi_2$ | 1.48 | 19.8 | 397.5 Ti-O-N | 1.63 | 28.83 | 99.35 $Si^0$ | 1.44 | 36.74 |
| | 455.89 Ti-O-N, Ti-O | 2.92 | 28.63 | 399.55 N-O (Si/Ti) | 2.32 | 8.75 | 102.29 Si-O-N/Si-O | 0.54 | 4.25 |
| **(650 °C )** | 454.74 TiN | 1.61 | 59.31 | 396.99 TiN | 1.33 | 52.97 | 98.78 $TiSi_2$ | 1.07 | 48.63 |
| | 454.01 $TiSi_2$ | 1.64 | 14.5 | 397.52 Ti-O-N | 1.15 | 33.03 | 99.47 $Si^0$ | 1.2 | 44.85 |
| | 456.12 Ti-O-N, Ti-O | 2.84 | 26.17 | 398.84 N-O (Si/Ti) | 1.83 | 14 | 97.78 | 0.7 | 6.52 |

(*Sum of Ti $2p_{3/2}$ and Ti $2p_{1/2}$)